\documentclass[journal = jctcce, manuscript=article]{achemso}

\usepackage{url}
\usepackage[english]{babel} 
\usepackage[utf8]{inputenc} 
\usepackage[T1]{fontenc}
\usepackage[normalem]{ulem}
\usepackage[hidelinks]{hyperref}

\usepackage{amssymb} 
\usepackage{mathalfa} 
\usepackage{algorithm} 
\usepackage{algpseudocode} 
\usepackage[table]{xcolor}
\usepackage{array}
\usepackage{titlesec}
\usepackage{mdframed}
\usepackage{listings}
\usepackage[section]{placeins}
\usepackage{caption}
\usepackage{subcaption}
\usepackage{braket}

\usepackage{amsmath} 
\usepackage{booktabs}

\newcolumntype{L}[1]{>{\raggedright\let\newline\\\arraybackslash\hspace{0pt}}m{#1}}
\newcolumntype{C}[1]{>{\centering\let\newline\\\arraybackslash\hspace{0pt}}m{#1}}
\newcolumntype{R}[1]{>{\raggedleft\let\newline\\\arraybackslash\hspace{0pt}}m{#1}}

\newcommand{\cn}{\color{black}}

\usepackage[nolist,nohyperlinks]{acronym}

\title{Tucker tensor approach for accelerating exchange computations in a real-space finite-element discretization of generalized Kohn-Sham density functional theory}
\author{Vishal Subramanian}
\affiliation{Department of Materials Science and Engineering, University of Michigan, Ann Arbor, Michigan 48109, USA}
\author{Sambit Das}
\affiliation{Department of Mechanical Engineering, University of Michigan, Ann Arbor, Michigan 48109, USA}
\author{Vikram Gavini}
\affiliation{Department of Mechanical Engineering, University of Michigan, Ann Arbor, Michigan 48109, USA}
\alsoaffiliation{Department of Materials Science and Engineering, University of Michigan, Ann Arbor, Michigan 48109, USA}
\email{vikramg@umich.edu}

\begin{document} 
\maketitle

\begin{abstract}
The evaluation of Fock exchange is often the computationally most expensive part of hybrid functional density functional theory calculations in a systematically improvable, complete basis. In this work, we employ a Tucker tensor based approach that substantially accelerates the evaluation of the action of Fock exchange by transforming 3-dimensional convolutional integrals into a tensor product of 1-dimensional convolution integrals. Our numerical implementation uses a parallelization strategy that balances the memory and communication bottlenecks, alongside overalapping compute and communication operations to enhance computational efficiency and parallel scalability. The accuracy and computational efficiency is demonstrated on various systems, including Pt clusters of various sizes and a $\text{TiO}_{\text{2}}$ cluster with 3,684 electrons. 

\begin{figure}[h]
    \centering
    \includegraphics[width=3.25in,height=1.75in]{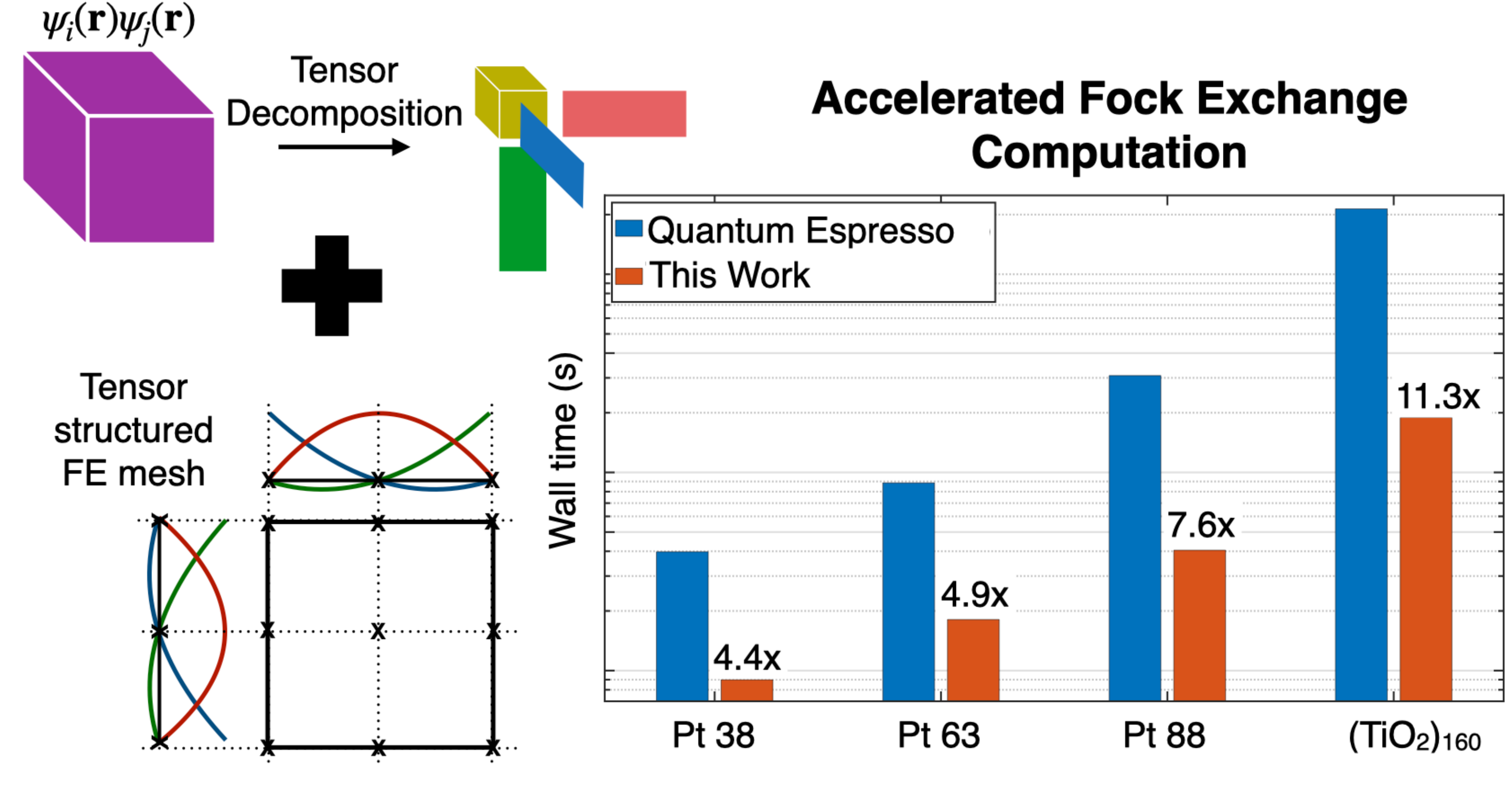}
    \caption{Graphical abstract}
    \label{fig:graphical_abs}
\end{figure}	

\end{abstract}

	\section{Introduction}
	\label{sec:dft}
	
Kohn-Sham Density functional theory~\cite{PhysRev.136.B864,PhysRev.140.A1133} (DFT) is widely used to predict various ground state properties of materials~\cite{martin2020electronic}. The wide adoption of DFT can be attributed to its balance between computational complexity and the accuracy obtained~\cite{hafner2006toward}. In principle, DFT is exact, however, it involves an unknown exchange-correlation (XC) functional of the electron density that encapsulates the many-body quantum-mechanical interactions. While the existence of the universal XC functional is known, its functional form is an open question to date~\cite{becke2014perspective}. The development of accurate and transferable exchange-correlation functionals continues to be an active research area~\cite{mardirossian2017thirty}. The progress in the past many decades has provided model XC functionals that have successfully predicted phase diagrams, mechanical, electronic, and magnetic properties of materials~\cite{martin2020electronic,kim2006molecular,jhi2000electronic}. Based on the nature of approximation used in the models, the XC functionals can be classified into different classes such as local density approximation (LDA) (dependent on local electron density)\cite{kohn1965self}, generalized gradient approximation (GGA) (dependent on local electron density and its gradient) \cite{PhysRevLett.77.3865} , hybrid functionals (also has a dependence on single-electron wavefunctions), etc. In general, studies have shown that the accuracy of the hybrid XC functionals, such as B3LYP \cite{becke98density}, PBE0\cite{pbe0Citation}, HSE06 \cite{doi:10.1063/1.1564060,krukau2006influence}, is better than that of LDA and GGA functionals when compared with experimental observations~\cite{finazzi2008excess,da2007hybrid,garza2016predicting}. In particular, the accuracy of the band gap calculation is significantly improved by hybrid functionals ~\cite{heyd2005energy,muscat2001prediction,salzner1997design,xiao2011accurate}. 
This improved accuracy of  hybrid functionals is attributed to including a fraction of the single-electron wavefunction-dependent Fock exchange energy. This inclusion of the wavefunction-dependent quantities has been formally justified under the generalized Kohn-Sham framework \cite{seidl1996generalized}.

The Fock exchange is a non-local operator. Its explicit construction scales as $\mathcal{O}({N_e}^4)$, where $N_e$ is the number of the electrons, as compared to the cubic-scaling cost of DFT calculations using semi-local functionals. Instead of the explicit construction, the action of exchange operator on an orbital can be computed by performing convolution integrals. The number of convolution integrals required at each iteration scales quadratically with system size, making it computationally expensive nevertheless. Many algorithms have been proposed to accelerate the exact exchange calculations. Broadly, these algorithms can be classified into two categories: (i) algorithms that reduce the number of times the action of the exchange operator needs to be computed; (ii) algorithms that reduce the computational complexity of the action of the exchange operator itself. Adaptively Compressed Exchange (ACE)~\cite{lin2016adaptively} is an example of the former, wherein a low-rank approximation for the exchange operator, which is exact in the space spanned by the occupied orbitals, is constructed. ACE approach has been shown to drastically reduce the cost of hybrid DFT calculations~\cite{lin2016adaptively}. However, even with ACE, the cost of a hybrid calculation can be substantially higher than using GGA, and the computational cost associated with the action of the exchange operator still remains a bottleneck. Hence, various reduced-order scaling strategies  such as the interpolative separable density fitting (ISDF) approach~\cite{hu2020accelerating,doi:10.1021/acs.jctc.7b00807}, and linear scaling methods like integral screening approaches and integral evaluation schemes~\cite{ko2020enabling,ko2021enabling,rettig2023even} have been proposed. ISDF approach exploits the rank deficiency in the product of the wavefunctions that appear in the convolution integrals, while linear-scaling methods utilize the locality of the wavefunctions to reduce the number of convolution integrals required. Though these approaches can reduce the computational cost, they do suffer from some shortcomings. Mainly, the robustness and systematic convergence of these algorithms are not well established. Further, the localization strategies employed in linear-scaling methods may not be generically applicable, especially for metallic systems. Various linear-scaling approaches also require the explicit construction of the exchange operator and hence are not amenable to complete basis-sets like plane-waves and finite-elements.

Over the years, many codes have been developed that solve the eigenvalue problem arising in the Kohn-Sham formalism, with atomic orbital type basis~\cite{apra2020nwchem,blum2009ab,frisch2009gaussian} or the plane-wave basis~\cite{giannozzi2009quantum,Gonze20092582,hafner2008ab} being popular discretization choices.
The atomic orbital basis is well-suited for systems such as molecules and atomic clusters. However, systematic numerical convergence with this basis, especially for metallic systems, is a concern. 
On the other hand, DFT codes based on the systematically convergent plane-wave basis exhibit limited parallel scalability due to the global nature of the basis. These limitations have restricted typical DFT calculations to systems containing a few thousand electrons. However, large-scale DFT simulations are useful in a wide range of scientific studies, such as understanding the energetics of defects, understanding the electronic structure of large biomolecules, and studying diffusivity in solids at interfaces, to name a few. This need for large-scale DFT calculations spurred the development of numerical methods that can handle large-scale problems efficiently. In particular, the recent use of a higher-order finite-element (FE) basis in electronic structure calculations~\cite{motamarri2013higher,pask2005finite} has shown significant promise. The FE basis is a piecewise polynomial basis with desirable features such as locality of the basis which is suitable for good parallel scalability, spatial adaptivity to efficiently handle non-periodic systems and all-electron calculations, and its ability to handle general boundary conditions (periodic, non-periodic and semi-periodic). DFT-FE~\cite{motamarri2020dft,das2022dft}---a massively parallel, open-source, real-space DFT code using adaptive higher-order FE discretization---has been shown to handle large-scale systems with $\mathcal{O}(10^{5})$ electrons, exhibiting good parallel scalability on many-core and hybrid CPU-GPU architectures \cite{das2019fast,GB2023}, and has recently been employed in studying a range of problems, including the electronic structure of DNA molecules~\cite{zhuravel2020backbone}, dislocations in crystalline materials~\cite{das2019fast,kumar2023effect}, phase transformations in doped nano-films~\cite{yao2022modulating}, computing spin Hamiltonian parameters of systems with defects~\cite{ghosh2019all,krishnendu2021spin}, and tackling the inverse DFT problem~\cite{kanungo2019exact,kanungo2021comparison,kanungo2023}.

In this work, we extend the capability of DFT-FE, presently limited to semi-local exchange-correlation functionals, to efficiently handle hybrid functionals. To this end, we build on the ideas presented in~\cite{ khoromskaia2015tensor,khoromskij2007low,khoromskij2009tensor} to implement a tensor-based algorithm to accelerate the computation of the action of the exchange operator. In particular, the product of single-particle wavefunctions appearing in the exchange operator is approximated using a low-rank Tucker tensor decomposition with systematic error control. Further, the $1/r$ kernel appearing in the exchange operator is approximated using a sum of Gaussians~\cite{braess2009efficient}. These systematically controllable approximations transform the 3-D convolution integrals into a tensor product of 1-D convolution integrals. This work presents many key numerical aspects of the implementation that enable an efficient and scalable treatment of the action of the exchange operator. We use an auxiliary tensor-structured finite-element mesh employing higher-order finite-elements for a systematically convergent and efficient evaluation of the convolution integrals in the application of the exchange operator, thereby significantly reducing the computational cost. In order to efficiently handle large-scale systems and obtain good parallel scalability, 
we have employed communication patterns that strike a balance between the memory required per MPI task and the amount of data communicated. In particular, the communication patterns have been designed to minimize data transfer and allow asynchronous communication and compute. Our numerical implementation also developed a fast interface for the transfer of fields between the unstructured finite-element mesh used to solve the Kohn-Sham eigenvalue problem and the tensor-structured finite-element mesh used for evaluating the action of the exchange operator. 

We demonstrate the accuracy, efficiency, and parallel scalability of the implementation by considering benchmark pseudopotential calculations on platinum (Pt) nanoclusters of increasing sizes, up to 1,584 electrons and a large $\text{TiO}_{\text{2}}$ cluster containing 3,648 electrons. The accuracy of computed ground-state energies has been ascertained by comparison with Quantum Espresso~\cite{giannozzi2017advanced}, a state-of-the-art open-source plane-wave code. In terms of computational efficiency, the present implementation is up to $\sim10\times$ faster than Quantum Espresso for the application of the exchange operator, which in turn results in considerable speed-ups for the solution time for the full ground-state problem. The scalability of the implementation is demonstrated on a 38 atom Pt nanocluster, with the code exhibiting good parallel scalability even at 128 nodes for a small system.

The remaining paper is structured as follows. Section \ref{ch:prelim} presents the mathematical background relevant to the DFT formulation for hybrid functionals and efficient solution procedures for the governing equations. Section \ref{ch:algo} presents the algorithmic details and numerical implementation of the action of exchange operator using the Tucker tensor based approach. Section \ref{ch:results} presents the results demonstrating the accuracy and performance of the algorithm and numerical implementation. Finally, we conclude with an outlook in Section~\ref{ch:conclusion}. 

	\section{Background}
	\label{ch:prelim}

 In density functional theory \cite{PhysRev.136.B864,levy1979universal}, the total electronic energy can be written as the sum of a universal functional of electron density and the energy due to interaction with an external potential. The ground state energy is determined by performing a minimization of the functional over the N-representable densities:
\begin{equation}
     E_{\textrm{ground-state}} = \min_{\rho(\mathbf{r})} \Big\{ \int V_{ext}(\mathbf{r}) \rho(\mathbf{r}) d\mathbf{r} + F[\rho]  \Big\}\,.
     \label{eq:gsEnergy}
 \end{equation}
  The universal functional F[$\rho$] is obtained by minimising the expectation of $\hat{T} + \hat{V}_{ee}$---where $\hat{T}$ is the many-body kinetic energy operator and  $ \hat{V}_{ee}$ is the many-body electron-electron interaction operator---over all anti-symmetric many-body wavefunctions that yield the density $\rho(\mathbf{r})$:
  \begin{equation}
     F[\rho] = \min_{\Psi \to \rho(\mathbf{r})} \{ \braket{\Psi| \hat{T} + \hat{V}_{ee} | \Psi} \}\,.
     \label{eq:hkt}
 \end{equation}
 In the generalised Kohn-Sham (GKS)~\cite{seidl1996generalized,garrick2020exact} approach, the many-body interacting system is replaced by an auxiliary system that can be represented by a single Slater determinant and has the same ground state density. Since the auxiliary system is represented by a single Slater determinant, it is described by single particle equations. By including energy functional $S$ that depends on the single Slater determinant($\mathbf{\{\phi\}}$), the auxiliary system can partly account for the quantum electron-electron interactions. The corresponding density functional ($F_S[\rho]$) is then given by
 \begin{equation}
     F_S[\rho] = \min_{\mathbf{\{\phi\}} \to \rho} S[\mathbf{\{\phi\}}]\,.
     \label{eq:Fs}
 \end{equation}
 The minimisation in Eq. \ref{eq:Fs} is performed over all single Slater determinants that yield the density $\rho$. Thus, the functional $F_S$ depends on the choice of $S[\mathbf{\{\phi\}}]$. The difference between $F[\rho]$ (defined in Eq.~\ref{eq:hkt}) and $F_S[\rho]$ (defined in Eq.~\ref{eq:Fs}) is denoted as the remainder functional $R_S[\rho]$. The exact form of $R_S[\rho]$ is unknown and is approximated in practice. 

 The $E_{\textrm{ground-state}}$ (Eq. \ref{eq:gsEnergy}) can thus be expressed as

 \begin{equation}
     E_{\textrm{ground-state}} = \min_{\rho(\mathbf{r})\to N} \Big\{ \int V_{ext}(\mathbf{r}) \rho(\mathbf{r}) d\mathbf{r} + F_S[\rho] + R_S[\rho]  \Big\}\,.
     \label{eq:gksEnergy}
 \end{equation}
Using the implicit dependence on the Slater determinant, the minimization in the above equation, reduces to a minimization over admissible Slater determinants leading to a non-linear eigenvalue problem commonly referred to as the generalized Kohn-Sham equation:
    	 \begin{equation}
     \Big( \hat{O} [\mathbf{\{\phi\}}]  + V_{ext}(\mathbf{r}) + V_{R}(\mathbf{r}) \Big) \psi_i (\mathbf{r}) = \epsilon_i \psi_i (\mathbf{r})\,,
     \label{eq:ksdft}
 \end{equation}
where $\hat{O} [\mathbf{\{\phi\}}] $ is a non-multiplicative potential obtained from the functional derivatives of $S[\mathbf{\{\phi\}}]$ with respect to the orbitals, and $V_{R}$ is a multiplicative potential obtained from the functional derivative of $R_S[\rho]$ with respect to $\rho$. In the case of hybrid functionals, $S[\mathbf{ \{\phi\}}]$ includes the non-interacting kinetic energy of electrons and a fraction of the exact exchange given by
\begin{equation}
S[\mathbf{\{\phi\}}] = \sum_i \int \psi_i(\mathbf{r}) \Big( -\frac{1}{2} {\nabla_i}^2 \Big) \psi_i (\mathbf{r}) d\mathbf{r}  - \frac{\alpha}{2} \sum_i \sum_j \int \int \frac{ \psi_i(\mathbf{r}) \psi_j(\mathbf{r}) \psi_j(\mathbf{r'}) \psi_i(\mathbf{r'}) }{|\mathbf{r} - \mathbf{r'}|}  d\mathbf{r'} d\mathbf{r}\,,
 \end{equation} 
 where $\alpha$ is generally chosen to be between $0$ and $1$. The above expressions are for a non-periodic system, which is the focus of this work, where the single-electron wavefunctions are real-valued. From the above, $\hat{O} [\mathbf{\{\phi\}}]$ is obtained to be
 \begin{equation}
     \hat{O} [\mathbf{\{\phi\}}]  = -\frac{1}{2} {\nabla_i}^2  + \alpha V_{\rm X} \,,
     \label{eq:ophi}
 \end{equation} 
where $V_{\rm X}$ denotes the exchange operator obtained from the functional derivative of the exact exchange. The exact form of remainder functional is unknown and an approximation often considered is given by $R_S[\rho]=E_{\rm H}[\rho]+(1-\alpha)E_{\rm X,sl}[\rho]+ E_{\rm c,sl}[\rho]$, where $E_{\rm H}$ denotes the Hartree energy, $E_{\rm X,sl}[\rho]$ and $E_{\rm c,sl}[\rho]$ denote the semi-local exchange and correlation parts, respectively, of model $\rho$-dependent exchange-correlation (XC) functionals. In hybrid XC functionals such as PBE0\cite{pbe0Citation}, semi-local functionals (PBE~\cite{PhysRevLett.77.3865}) are combined with a fraction of the exact exchange. The functional derivative of the above form of $R_S[\rho]$ is given by
 \begin{equation}
     V_R[\rho](\mathbf{r}) = V_{\rm H}[\rho](\mathbf{r}) + (1-\alpha) V_{\rm X,sl}[\rho](\mathbf{r}) + V_{\rm c,sl}[\rho](\mathbf{r})\,.
     \label{eq:vr}
 \end{equation} 
Finally, substituting the above terms in Eq.~\ref{eq:ksdft}, the following form of the GKS equation employed in hybrid DFT calculations is obtained:
\begin{equation}
     \Big( -\frac{1}{2} {\nabla_i}^2 + V_{ext}(\mathbf{r}) + V_H[\rho](\mathbf{r}) + \alpha V_{X} + (1-\alpha) V_{X,SL}[\rho](\mathbf{r}) + V_{c,SL}[\rho](\mathbf{r}) \Big) \psi_i (\mathbf{r}) = \epsilon_i \psi_i (\mathbf{r})\,.
     \label{eq:hybridxc}
\end{equation}

\subsection{Exchange operator}
The exchange operator is a non-local operator which requires the single-particle wavefunctions explicitly:  
 \begin{equation}
     (V_{\rm X})(\mathbf{r} ,\mathbf{r'} )  = - \sum_{j=1}^{N_e}  \frac{\psi_j(\mathbf{r}) \psi_j(\mathbf{r'})}{|\mathbf{r} - \mathbf{r'}|} \,.
     \label{eq:Exchange}
 \end{equation}
The explicit evaluation of the exchange operator  can be computationally expensive as the complexity scales as $ \mathcal{O}(N_e^4)$, where $N_e$ is the number of electrons in the system. Algorithms such as the resolution of identity~\cite{ren2012resolution,levchenko2015hybrid} have been proposed to accelerate this computation. However, these algorithms are not compatible with a complete basis as they are still prohibitively expensive.  The explicit construction of the exchange operator can be avoided by employing iterative algorithms to compute the action of the exchange operator on an occupied orbital $\psi_i$, and can be expressed as
 \begin{equation}
     (V_{\rm X}) \psi_i (\mathbf{r}) = - \sum_{j=1}^{N_e} \psi_j(\mathbf{r}) \int \frac{\psi_j(\mathbf{r'}) \psi_i(\mathbf{r'})}{|\mathbf{r} - \mathbf{r'}|} d\mathbf{r'}\,.
     \label{eq:ExchangeOperator}
 \end{equation}
Conventionally, numerical implementations based on systematically improvable and complete basis sets evaluate the convolution integrals arising in the action of the exchange operator by either solving a Poisson equation (method of choice for real-space basis sets) or using fast Fourier transforms (method of choice for plane-wave basis). The exchange operator requires occupied orbitals, which are obtained as the solution of the non-linear eigenvalue problem (cf. \eqref{eq:hybridxc}). Thus, these equations are often solved self-consistently, using a nested two-level self-consistent field (SCF) procedure as discussed below.

\subsection{Two-level nested SCF}
Two-level nested SCF refers to a nested fixed point iteration scheme involving inner and outer fixed point iterations. In the outer fixed point iteration, the wavefunction dependent parts of the Hamiltonian (exchange operator) are updated. In the inner SCF, the wavefunction dependent parts of Hamiltonian are held fixed and the electron-density dependent non-linear eigenvalue problem is solved until convergence. With such a split, robust mixing strategies such as Anderson mixing~\cite{anderson1965iterative} can be used to accelerate the convergence of the electron-density in the inner SCF iterations.
Within DFT-FE, we use a Chebyshev-filtered subspace iteration technique~\cite{zhou2006self,motamarri2013higher} to solve for the lowest occupied eigensubspace and the ground-state electron density. The Kohn-Sham (KS) orbitals obtained at the end of the inner SCF iterations are used to update the exchange operator.  The simulation is considered converged when the difference in exact-exchange energy between successive outer SCF iterations is below a pre-determined tolerance. Such a two-level nested SCF scheme is implemented in conjunction with Adaptively Compressed Exchange (ACE)~\cite{lin2016adaptively, hu2017adaptively} operator to reduce the computational cost further, which is discussed below.

 \subsection{ACE implementation}\label{sec:ace}
 The ACE implementation aims to construct a low-rank representation of the exchange operator~\cite{lin2016adaptively}. The exchange operator is generally full-rank, and a low-rank approximation can lead to aproximation errors. However, a low-rank approximation of the exchange operator that is exact for the occupied KS orbitals can be constructed and is the main idea behind the ACE implementation. The ACE operator, denoted by $V^{ACE}_{\rm X}$, is constructed such that it satisfies the Hermiticity and negative semi-definite nature of $V_{\rm X}$ 
  \begin{equation}
     V^{ACE}_{\rm X} (\mathbf{r} , \mathbf{r'}) = - \sum_{k=1}^{N_e}\xi_k (\mathbf{r})\xi_k (\mathbf{r'})\,,
  \end{equation}
where $\mathbf{\xi}_k $ is the projection vector defined as
  \begin{equation}
      \xi_k (\mathbf{r}) = \sum_{i=1}^{N_e} W_i (\mathbf{r}) \big(\mathbf{L}^{-1}\big)_{ki}\,.
      \label{eq:eta}
  \end{equation}
\cn
In the above, $W_i$ is the vector obtained from the action of the exchange operator on the $i^{th}$ KS orbital and is given by
\begin{equation}
      W_i (\mathbf{r}) = V_X[\psi(\mathbf{r})] \psi_i(\mathbf{r})\,,
      \label{eq:ace1}
  \end{equation}
and the matrix $\mathbf{L}$  is obtained by performing a Cholesky decomposition on matrix $\mathbf{M}$:
\begin{equation}
      \mathbf{M} = - \mathbf{L} \mathbf{L} ^T\,, \quad\quad M_{ij} = \int \psi_i(\mathbf{r}) W_j(\mathbf{r}) d\mathbf{r} \,.
\end{equation} 

\begin{figure}[h]
    \centering
    \includegraphics[width=0.7\textwidth]{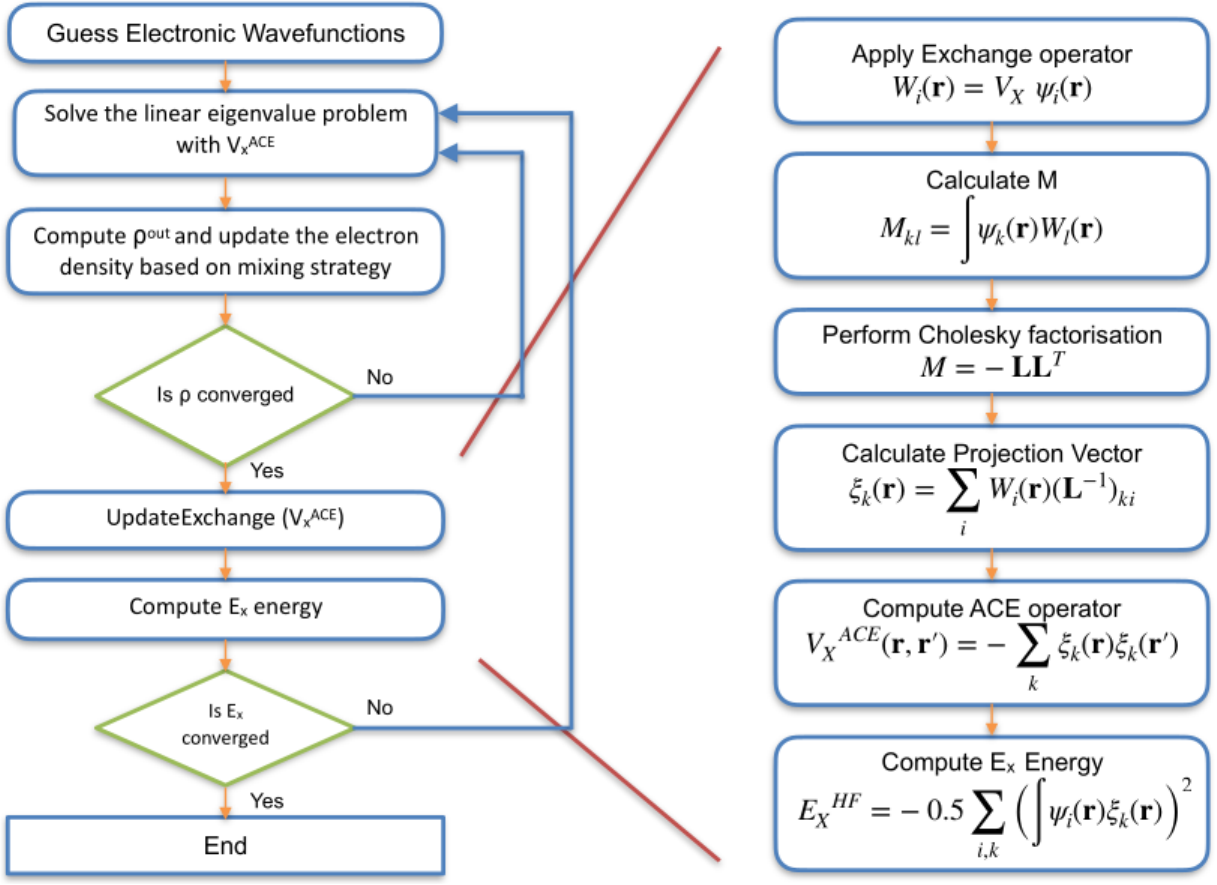}
    \caption{Nested SCF loop algorithm for hybrid DFT }
    \label{fig:flow}
\end{figure}

The flowchart of the entire algorithm for solving the GKS equations in hybrid DFT is depicted in Fig.~\ref{fig:flow}. The construction of the ACE operator, as described above, is performed in \texttt{updateExchange}.
The advantage of using ACE operator is that, once constructed, it significantly reduces the cost of action of the exchange operator on trial KS orbitals, thereby reducing the walltimes of the inner SCF iterations. The aim of this work is to reduce the cost of the calculations performed in \texttt{updateExchange} using Tucker-tensor decomposition.

\subsection{Tucker-tensor decomposition}

Tensor-structured techniques, in particular the Tucker-tensor decomposition, have been used to construct low-rank approximations of data and fields in three and higher dimensions---we refer to~\cite{kolda2009tensor} for a comprehensive review of the subject. The Tucker tensor decomposition of a field in 3-D is given by
\begin{equation}
    \phi(\mathbf{r}) \approx \sum_{\alpha=1}^{r1} \sum_{\beta=1}^{r2} \sum_{\gamma = 1}^{r3} \mathcal{G}_{\alpha \beta \gamma}  {f^\alpha}({x}){g^\beta}({y}) {h^\gamma}({z})\,.
    \label{eq:tucker}
 \end{equation}
In the above, $r=\max\{r_1,r_2,r_2\}$ denotes the rank of the Tucker decomposition, $\mathcal{G}_{\alpha \beta \gamma}$ denotes the core-tensor and $f^{\alpha}$, $g^{\beta}$, $h^{\gamma}$ are 1-D functions. Notably, the approximation error for regular fields in the Tucker decomposition decays exponentially with increasing Tucker rank---also demonstrated in Sec.~\ref{ch:convolutions}. For electronic structure, including single-particle wavefunctions and electron density, a Tucker rank of $\mathcal{O}(10)$ provides a good approximation to use in practical calculations~\cite{Khoromskij2009,Blesgen2012,Lin2021}.  

In the context of the present work, which seeks to accelerate the evaluation of the action of the Fock exchange operator on single-particle KS wavefunctions, the Tucker decomposition of the KS wavefunctions presents a path to reduce the convolution integrals in 3D to convolutions in 1D as discussed in Sec.~\ref{ch:convolutions}. A proof-of-concept demonstration of this idea was presented in~\cite{Khoromskij2009}, and this idea was employed in evaluating the convolution integrals arising in the computation of Hartree potential in a Tucker tensor basis~\cite{Motamarri2016,Lin2021}. In the present work, we build on these earlier efforts and develop an efficient and scalable implementation for hybrid DFT calculations in a systematically convergent and complete finite-element basis---a piecewise polynomial basis in real-space. In the following section we discuss the various aspects of the computational framework, the details of the numerical implementation, including the strategies for achieving an efficient and scalable implementation.   
\section{Algorithmic implementation}
\label{ch:algo}
In this section, we present the various aspects of our Tucker-tensor algorithm used to accelerate the action of the  Exchange operator on the Kohn-Sham orbitals.  
 
\subsection{Overview of the algorithm}
We provide a brief overview of the various steps in the Tucker-tensor algorithm before discussing the implementation details. To begin, we define a finite-dimensional subspace  ${V}^M_1$ spanned by an unstructured finite-element basis onto which we project the KS Hamiltonian for solving the inner fixed-point iteration on electron-density. Next, the KS wavefunctions ($\psi_i$) from the solution of the inner SCF iterations, belonging to ${V}^M_1$, are transferred to another finite-element basis which has a tensor structure, as this is required to perform Tucker-tensor decomposition. We refer to this subspace as ${V}^M_2$.
Subsequently, we compute the Tucker decomposition and the  convolution integrals in ${V}^M_2$, using the tensor structured basis. Finally, we compute action of the exchange operator $W_i=V_X \psi_i$ in ${V}^M_2$, and this is transferred back to ${V}^M_1$ as $\Tilde{W_i} \in {V}^M_1$. These steps are mathematically represented as:
\begin{subequations}
    \begin{equation} 
     \psi^{{V}^M_2}_i = T_{{V}^M_1 \to {V}^M_2}\ \ \psi^{{V}^M_1}_i \quad \forall \,i
     \label{eq:tranferUtoS}
    \end{equation}
    \begin{equation} 
     \varphi_{ji}^{{V}^M_2} (\mathbf{r}) = \int \frac{{\psi_j}^{{V}^M_2}(\mathbf{r'}){\psi_i}^{{V}^M_2}(\mathbf{r'})}{ | \mathbf{r} - \mathbf{r'} |} d \mathbf{r'} \quad \forall \,(i,j)
     \label{eq:conv}
    \end{equation}
    \begin{equation} 
    {W_i}^{{V}^M_2}  = \sum_j {\psi_j}^{{V}^M_2} \varphi_{ji}^{{V}^M_2} \quad \forall \,i
     \label{eq:hadam}
    \end{equation}
    \begin{equation} 
    {\Tilde{W_i}}^{{V}^M_1} = T_{{V}^M_2 \to {V}^M_1} {W_i}^{{V}^M_2} \quad \forall \,i
    \end{equation}
\end{subequations}
where $\psi_i$ refers to the wavefunctions, $T$ represents the transfer operation from one subspace to another, and $\varphi_{ji}$ represents the output of the convolution. The implementation details are discussed in the subsequent subsections.

As the subspaces ${V}^M_1$ and ${V}^M_2$ can be different in general, this can lead to a loss in accuracy during these transfers. To mitigate this issue, a structured finite-element (FE) basis with a commensurate mesh refinement as that of the unstructured FE basis is used. Using an FE basis for the Tucker-tensor decomposition provides a simplified, efficient, and accurate numerical framework to perform the interpolations required for the transfer and the evaluation of convolution integrals.

 \subsection{Creation of Structured FE basis}
\begin{figure}[h]
    \centering
    \includegraphics[width=0.7\textwidth]{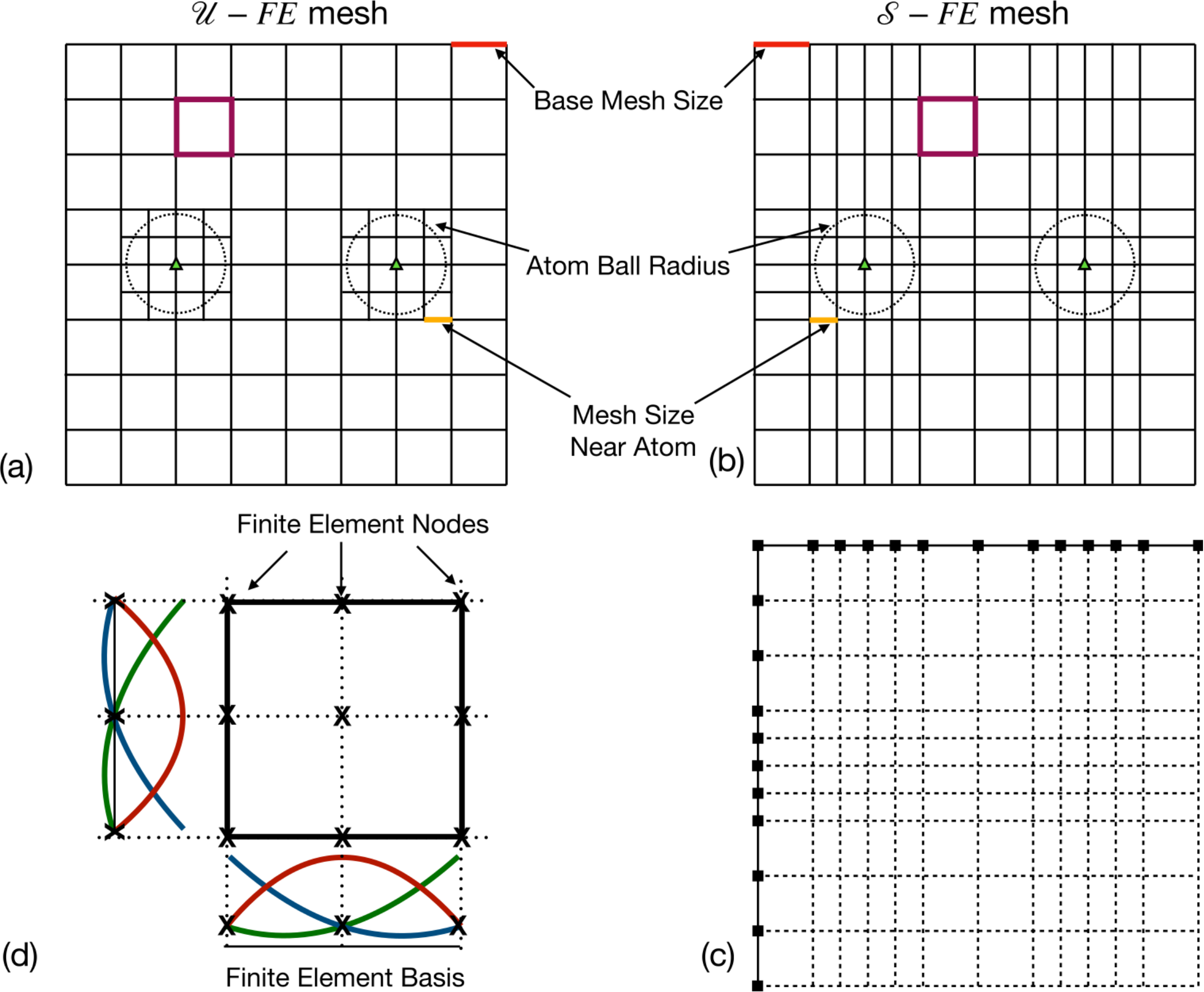}
    \caption{Schematic describing various aspects of the finite element meshes used in this work. (a)-(b) An unstructured finite element mesh ($\mathcal{U}-{\rm FE}$ mesh) and a structured finite element mesh ($\mathcal{S}-{\rm FE}$ mesh). In (a)-(b), the green triangle represents an atom. Selective refinement around the atom results in a non-uniform, adaptively refined mesh. The mesh parameters---Base Mesh Size, Atom Ball Radius, and Mesh Size Near Atom---are used to construct the spatially adaptive meshes. The purple lines mark the boundary of a finite element cell; (c) The structured mesh represented as a tensor product of 1-D meshes; (d) Finite element cell, with `x' markers denoting the finite element nodes. The finite element basis is the tensor product of the 1D Lagrange polynomials constructed from the finite element nodes.}
    \label{fig:schematic_mesh}
\end{figure}

Tensor structured operations require a structured mesh, whereas the FE basis for the Kohn-Sham problem can work in general with an unstructured mesh (henceforth denoting the corresponding mesh and basis as $\mathcal{U}-{\rm FE}$ mesh and basis). Hence, we create a structured mesh---denoting the corresponding mesh and basis as $\mathcal{S}-{\rm FE}$ mesh and basis---to aid the tensor-structured computations. To this end, we construct three 1-D meshes (one along each direction), and the tensor product of these 1-D meshes provides the desired tensor-structured 3-D mesh. 
Figure~\ref{fig:schematic_mesh} shows the schematic of the unstructured and structured meshes, the tensor structured nature of the $\mathcal{S}-{\rm FE}$ mesh, and the various aspects of FE discretization including FE cells, FE nodes and FE basis. We refer the readers to~\cite{hughes2012finite} for a more comprehensive discussion on the approximation properties of finite element discretization, and to~\cite{pask2005finite,motamarri2013higher,motamarri2020dft,tsuchida1995electronic,tsuchida1996adaptive} for the use of finite elements in electronic structure calculations.

The $\mathcal{S}-{\rm FE}$ mesh is generated by using similar mesh parameters as that of the $\mathcal{U}-{\rm FE}$ mesh, i.e. similar `base mesh-size' ($bms$; the initial uniform mesh size  before any adaptive refinement) and similar `mesh-size near atom' ($msna$; the smaller mesh-size near the nuclei, obtained by adaptive refinement). This is achieved by starting with a 1-D mesh set to the base mesh-size ($bms$), followed by selective refinement until the finite-element cells near an atom are smaller than the mesh-size near the atom ($msna$). The selective refinement uses an adaptive algorithm that refines finite-element cells based on their proximity to an atom. The cells are then populated with FE nodes. 
 
In order to reduce the peak memory requirements, the coordinates of the FE nodes in the $\mathcal{S}-{\rm FE}$ mesh are not stored. Instead, the coordinates of the FE nodes in the 1-D meshes are stored, from which the coordinates of the 3-D mesh are computed on the fly. We employ a lexicographic numbering scheme that simplifies the map from nodal coordinates in the 3-D mesh to the nodal coordinates in the three 1-D meshes, and vice versa. 
 
\subsection{Transfer of wavefunctions between Unstructured and Structured FE basis}
The first step in the Tucker-tensor algorithm to accelerate the action of the exchange operator is the transfer of the  wavefunctions in the $\mathcal{U}-{\rm FE}$ basis to the $\mathcal{S}-{\rm FE}$ basis (cf. Eq. \ref{eq:tranferUtoS}), which is performed by interpolation. In the FE basis, interpolation to an arbitrary point requires the evaluation of the shape functions---basis functions local to a given FE cell---at the point of interest. 
To this end, corresponding to each $\mathcal{U}-{\rm FE}$ cell, we find all the FE nodes of the $\mathcal{S}-{\rm FE}$ mesh that lie in the cell. Subsequently, the relevant $\mathcal{U}-{\rm FE}$ shape function values at the $\mathcal{S}-{\rm FE}$ node locations are evaluated. This map and the shape function values, used for the interpolation, are computed once and stored \emph{a priori}.      
 
Using the stored shaped function values, we now discuss a cache-efficient implementation to perform the transfer operation. In the $\mathcal{U}-{\rm FE}$ basis, we store the values of all KS wavefunctions for all nodes in a contiguous memory layout with the wavefunction index for a given FE node being the fastest index in this storage. Since the shape functions are independent of the wavefunctions, the interpolation from one cell of the $\mathcal{U}-{\rm FE}$ to the relevant nodes of the $\mathcal{S}-{\rm FE}$ is performed simultaneously for all the wavefunctions. This can be recast as the following matrix-matrix multiplication for each $\mathcal{U}-{\rm FE}$ cell
 \begin{equation}
{\psi^{\mathcal{S}-FE}_{i,a}} = \sum_b N_{ab} {\psi^{\mathcal{U}-FE}_{i,b}}\,,
\label{eq:interpolate}
\end{equation}
where $a$ and $b$ refers to the nodal index in the $\mathcal{S}-{\rm FE}$ and $\mathcal{U}-{\rm FE}$ basis, respectively. $N_{ab}$ refers to the shape function value of the $b^{th}$ node of the $\mathcal{U}-{\rm FE}$ basis at the spatial location of the $a^{th}$ nodal point of $\mathcal{S}-{\rm FE}$ basis.  

The transfer of fields computed in the $\mathcal{S}-{\rm FE}$ basis (cf. Eqs.~\eqref{eq:conv}) and \eqref{eq:hadam}) to the $\mathcal{U}-{\rm FE}$ basis proceeds along similar lines. 

\subsection{Convolution operation}\label{ch:convolutions}
\begin{figure}[htbp!]
    \centering
    \includegraphics[width=0.7\textwidth]{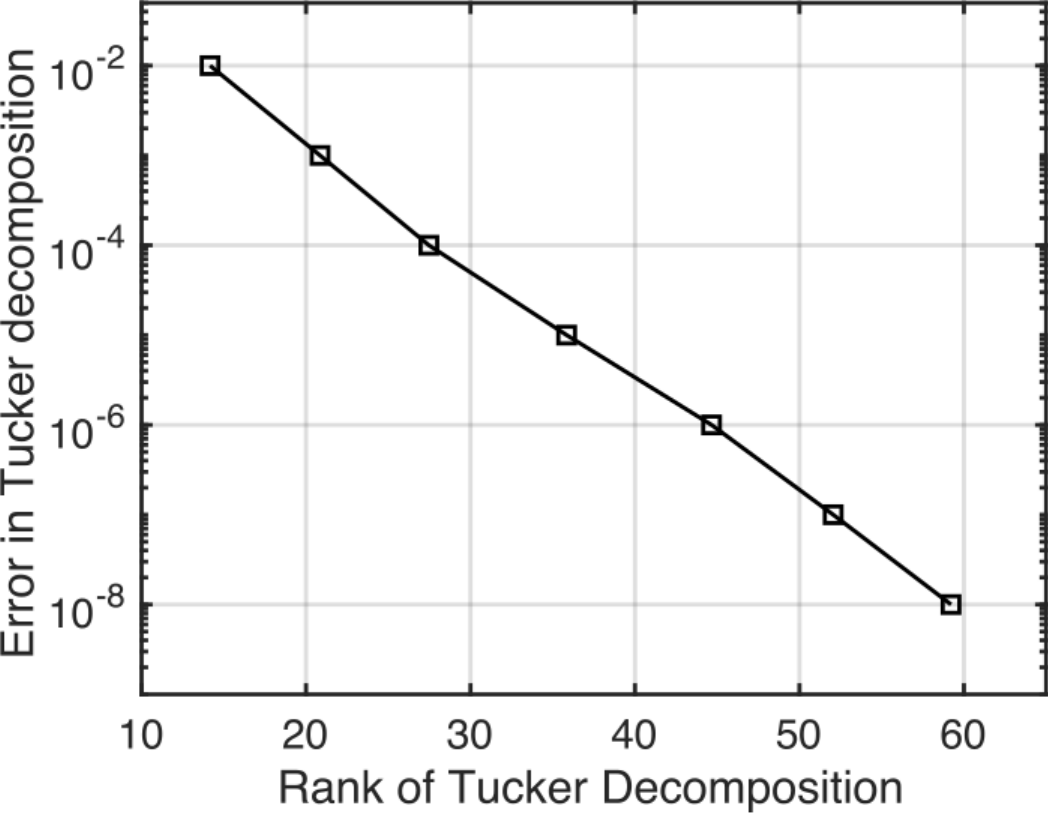}
    \caption{Approximation error in the Tucker decomposition of $\psi_j (\mathbf{r'} ) \psi_i (\mathbf{r'} )$, measured in terms of the relative error in Frobenius norm on the FE nodes as a function of Tucker rank. Benchmark system: Pt-14 cluster.}
    \label{fig:rankTol}
\end{figure}

In order to compute 
\begin{equation}
     \varphi_{ji}(\mathbf{r})  \equiv \int \frac{\psi_j (\mathbf{r'} ) \psi_i (\mathbf{r'} )}{ | \mathbf{r} - \mathbf{r'} |}d\mathbf{r'} \,,
     \label{eq:varphi}
 \end{equation}
 we compute $\psi_j (\mathbf{r'} ) \psi_i (\mathbf{r'} )$ by trivially multiplying the fields at the FE nodes of $\mathcal{S}-{\rm FE}$ mesh.
Subsequently, we perform a Tucker decomposition for every pair of $i$-$j$ wavefunctions using the STHOSVD  algorithm implemented in TuckerMPI ~\cite{ballard2020tuckermpi}, providing the low-rank approximation 
    \begin{equation}
     \psi_j (\mathbf{r'} ) \psi_i (\mathbf{r'} ) \approx \sum_\alpha \sum_\beta \sum_\gamma  \mathcal{G}_{\alpha \beta \gamma} f_{ij}^\alpha(x') g_{ij}^\beta(y')  h_{ij}^\gamma(z')\,.
     \label{eq:TuckerDecompField}
 \end{equation}
The error in the Tucker decomposition decreases exponentially with the Tucker rank as depicted in Fig.~\ref{fig:rankTol} for a 14-atom Pt nanocluster, a benchmark system used in this study. Notably, very low approximation errors can be achieved with modest values of rank around 50.
 
Further, to enable a 1-D separable form for fast evaluation of convolution integrals, we approximate the $1/r$ kernel using a canonical tensor decomposition with 1-D Gaussian functions~\cite{braess2009efficient} as
   \begin{equation}
     \frac{1}{|\mathbf{r} - \mathbf{r'}|} \approx \sum_{t=1}^T \mu_t exp(-\theta_t (x - x')^2) exp(-\theta_t (y - y')^2) exp(-\theta_t (z - z')^2)\,.
     \label{eq:gauss}
 \end{equation}
 \begin{figure}[h]
    \centering
    \includegraphics[width=0.7\textwidth]{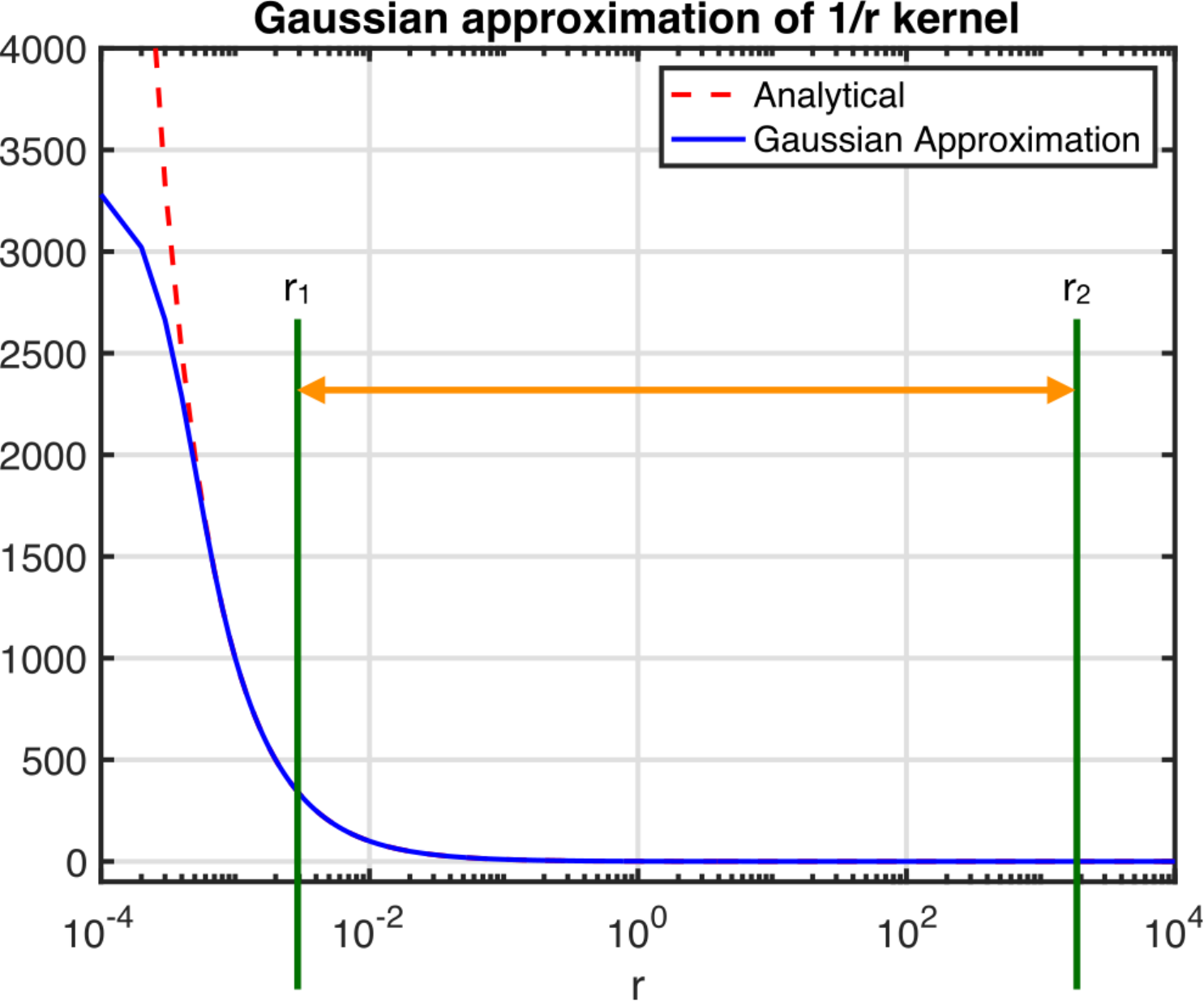}
    \caption{Accuracy of the Gaussian approximation for the $1/r$ kernel with T=35.}
    \label{fig:gaussApprox}
\end{figure}
 The accuracy of this approximation for the $1/r$ kernel is depicted in Fig. \ref{fig:gaussApprox}. Though the approximation fails to capture the asymptotical behavior of the kernel accurately at small $r$, the accuracy of the convolution integral is negligibly affected as long as the error in the approximation is controlled within a region $[{r_1},{r_2}]$. The smallest spacing between the 1-D FE nodes and the domain size determines the bounds ${r_1}$ and ${r_2}$, respectively. 
Finally, using Eqs.~\ref{eq:TuckerDecompField} and \ref{eq:gauss} in Eq.~\ref{eq:varphi}, the 3-D convolution integrals can be well approximated as a tensor product of 1-D convolution integrals as
\begin{equation} \label{eq:final}
\begin{split}
\varphi_{ji}(\mathbf{r})  \approx & \sum_{\alpha \beta \gamma} \int \int \int  \sum_t   \Big\{ \mathcal{G}_{\alpha\beta\gamma} f_{ij}^\alpha(x') g_{ij}^\beta(y')  h_{ij}^\gamma(z') \\  & \mu_t exp(-\theta_t (x - x')^2) exp(-\theta_t (y - y')^2) exp(-\theta_t (z - z')^2) \Big\} dx' dy' dz' \,,
\\ 
   = & \sum_t \sum_{\alpha \beta \gamma}  \mathcal{G}_{\alpha \beta \gamma}  \mu_t  \Big[  \int f_{ij}^\alpha(x')  exp(-\theta_t (x - x')^2)   dx'\Big]\times \\  & \Big[\int g_{ij}^\beta(y')  exp(-\theta_t (y - y')^2) dy' \Big] \times \Big[ \int h_{ij}^\gamma(z')  exp(-\theta_t (z - z')^2) dz'\Big]\,. 
\end{split}
\end{equation}

The evaluation of Eq.~\ref{eq:final} requires a numerical evaluation of the 1-D integrals, which are computed using the 1-D FE meshes that generate the $\mathcal{S}-{\rm FE}$ mesh (cf. Figure~\ref{fig:schematic_mesh}(c)). Numerical integration using the 1-D FE basis involves interpolating to a set of quadrature points and summing with appropriate quadrature weights. The number of quadrature points required to integrate the 1-D functions in Eq.\eqref{eq:final} accurately can be high due to the sharply varying  exponentials in the integrand. However, the number and location of quadrature points and the powers of the exponentials are all known \emph{a priori}. Since these values are independent of the wavefunction values, they can be precomputed. Once these values are precomputed, the numerical integration of the 1-D integrals can be recast as a summation over FE nodal values of wavefunction specific quantities and the precomputed values. As the number of nodal points are much lower than the number of quadrature points, a significant reduction in computation cost can be obtained. Mathematically, the recasting of the compute operations can be expressed as follows: the interpolation of the field $ f_{ij}^\alpha(x')$ to the quadrature points is a weighted sum of the nodal values given by  
 \begin{equation}
 \label{eq:quadVal}
   f_{ij,b}^{\alpha} = \sum_a \xi_{ab} f_{ij,a}^{\alpha}\,,
 \end{equation}
where $\xi_{ab}$ is the 1-D shape function of the $a^{th}$ node at the $b^{th}$ quadrature point. Further, the 1-D convolution integrals numerically evaluated at any location $x$ are given by 
  \begin{equation}
  \label{eq:discInteg}
  \int f_{ij}^{\alpha}(x')  exp(-\theta_t (x - x')^2)   dx' = \sum_b
  f_{ij,b}^{\alpha} exp(-\theta_t (x - x_b)^2) Jw_b \,,
 \end{equation}
where the summation is over the quadrature points $b$, $x$ represents the spatial location at which the convolution integrals are evaluated, and $Jw_b$ represents the determinant of the 1-D finite-element Jacobian times the quadrature weight. Using the expression for $f_{ij,b}^{\alpha}$ from  Eq.~\ref{eq:quadVal} in Eq.~\ref{eq:discInteg}, we get
  \begin{equation}
  \int f_{ij}^\alpha(x')  exp(-\theta_t (x - x')^2)   dx' = \sum_b  \Big(\sum_a \xi_{ab} f_{ij,a}^\alpha \Big) exp(-\theta_t (x - x_b)^2)  Jw_b \,.
 \end{equation}
Rearranging the summation, we obtain
   \begin{equation}
  \int f_{ij}^\alpha(x')  exp(-\theta_t (x - x')^2)   dx' = \sum_a \Big( \sum_b  exp(-\theta_t (x - x_b)^2) Jw_b \xi_{ab} \Big) f_{ij,a}^\alpha\,,
 \end{equation}
where the summation over $b$ (quadrature points) can be precomputed once for the 1-D FE basis and stored. The summation over $a$ involves wavefunction specific values ($f_{ij,a}^\alpha$), which has to be performed repeatedly, but the computational cost of this summation is cheap as the number of FE nodes in 1-D mesh are $\mathcal{O}(10)-\mathcal{O}(100)$.

The convolution integrals are evaluated at the nodal locations of the 1-D mesh, i.e $x$ is set to the spatial location of FE nodes in 1-D FE mesh. Hence, the above summation has to be performed for all pairs of $i$-$j$, at all nodes FE nodes in the 1-D mesh, for each $t$ along each direction. These computations are performed as matrix-matrix products by choosing appropriate memory storage for all the above variables. Once the convolution of every $i$-$j$ pair is computed, we reconstruct the full 3-D tensor and multiply with $\psi_j$ (cf.~Eq.~\ref{eq:hadam}). Finally, the sum along the $j$ index provides the action of the exchange operator on $\psi_i$, computed in $\mathcal{S}-FE$ basis. In our implementation, this step of convolution operations---the evaluation of the 1-D convolution integrals---together with reconstruction, forms the most computationally expensive part of the calculation.
 
\subsection{MPI implementation} 
\begin{figure}[h]
    \centering
    \includegraphics[width=0.7\textwidth]{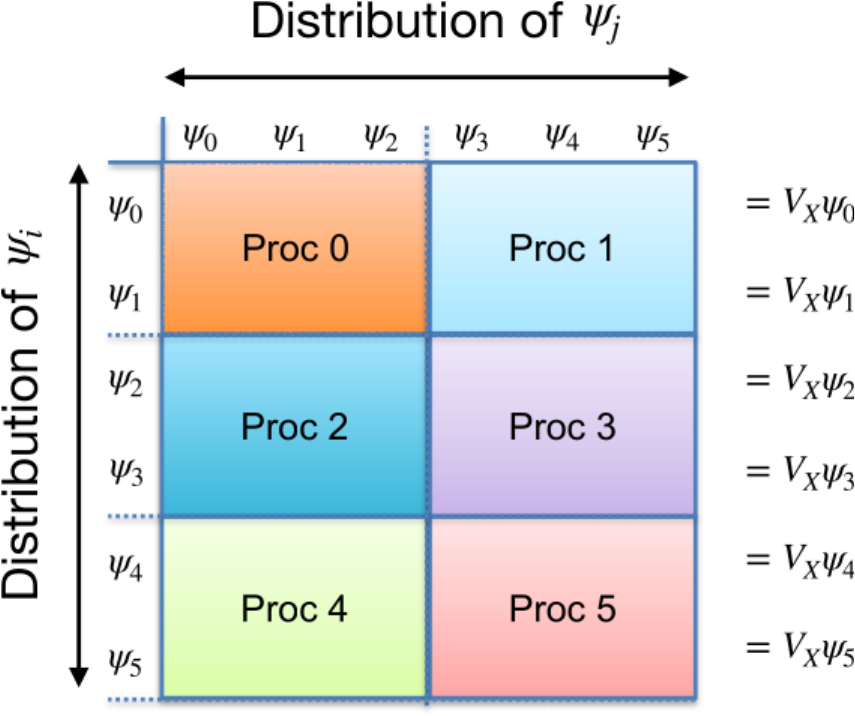}
    \caption{MPI task parallelization for KS wavefunctions in the structured FE mesh, shown for six wavefunctions distributed over six tasks. Summing the output from the tasks along a row provides the action of the exchange operator.}
    \label{fig:taskMpiComm}
\end{figure}

\begin{figure}[h]
    \centering
    \includegraphics[width=0.7\textwidth]{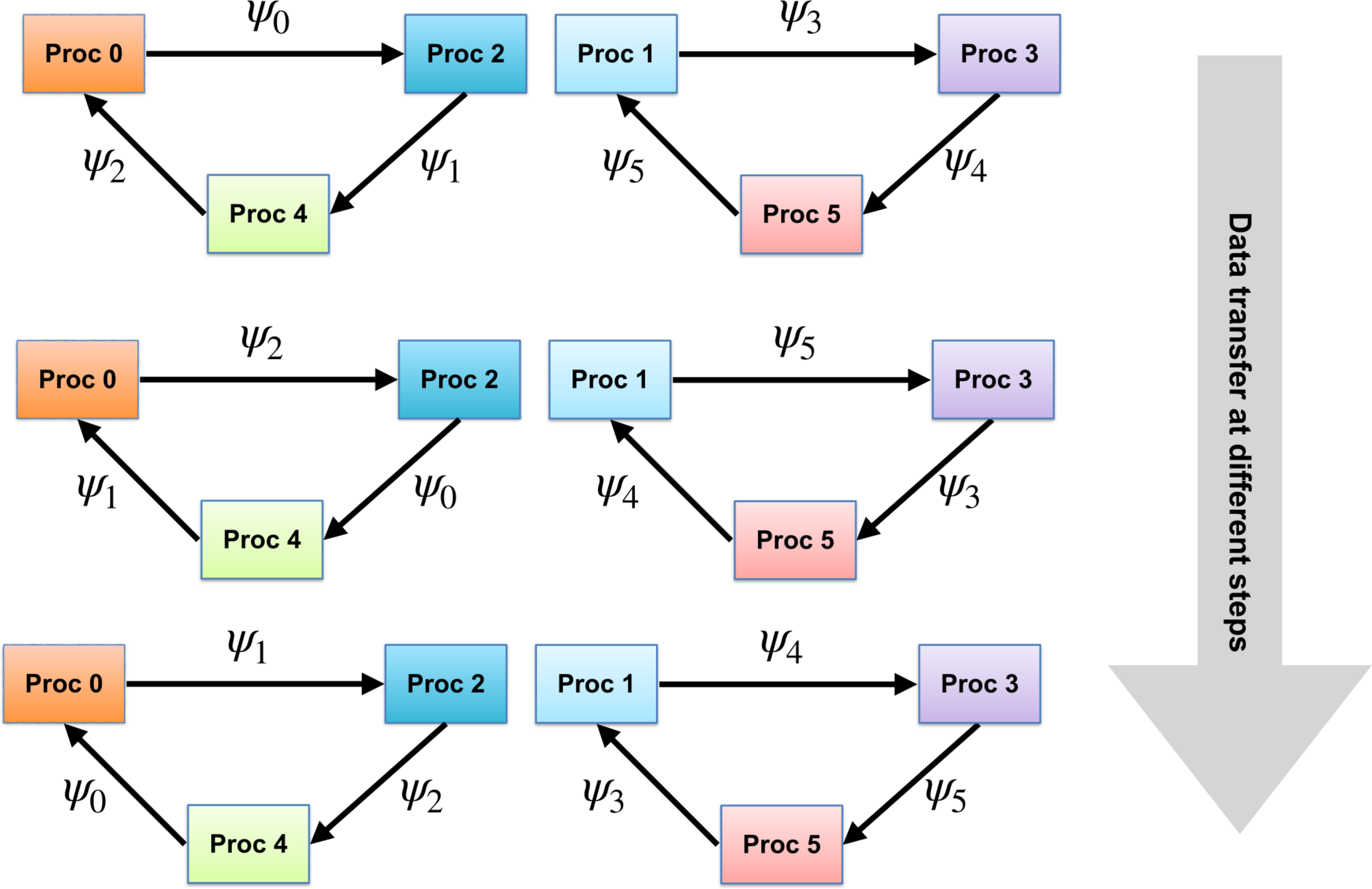}
    \caption{Representation of memory parallelization for different subgroups of MPI tasks using round-robin algorithm. At the end of each step, the wavefunction is transferred to the next task using MPI point-to-point communication calls.}
    \label{fig:memMpiComm}
\end{figure}

In order to enable large-scale computations involving thousands of electrons, a scalable numerical implementation of the Tucker-tensor algorithm is crucial. To this end, we now discuss our task/memory parallelization and MPI communication strategy. Eq. \ref{eq:varphi} requires the computation of $N_e^2$ convolutions, corresponding to each $i$-$j$ pair. A high level of task parallelization can be achieved by distributing these $N_e^2$ convolutions equitably among the tasks. For the purpose of computing the convolution integrals, it is efficient to have the KS wavefunction fields whose convolution are being computed to be local to the task. Thus, an efficient layout for storing KS wavefunctions in the $\mathcal{S}-{\rm FE}$ basis is band parallelism over the processor grid as shown in Fig.~\ref{fig:taskMpiComm}. To elaborate, each task computing a subset of the convolution integrals has the relevant KS wavefunctions in the $\mathcal{S}-{\rm FE}$ basis local to the processor. The column index range for each MPI task determines the subset of $j$-indexed wavefunctions assigned to the task. Similarly, the task's row index range for each MPI task determines the subset of $i$-indexed wavefunctions assigned to  it. Upon computing the convolution integrals, summing the output obtained from the tasks along a row of the processor grid provides the final output of the action of the exchange operator on a wavefunction $\psi_i$ ($V_X \psi_i$). We note that such a memory parallelization layout for KS wavefunctions in $\mathcal{S}-{\rm FE}$ basis is efficient from a communication standpoint, as no transfer of wavefunctions between tasks is necessary while computing the convolution integrals. However, this strategy increases the memory footprint as duplicate copies of KS wavefunctions are stored---for instance, in the layout in Fig.~\ref{fig:taskMpiComm}, $\psi_0$ is stored in Proc~0, Proc~1, Proc~2 and Proc~4. While this does not pose a limitation on large number of tasks, the increased memory footprint can be an issue when running the simulation on a limited number of nodes. To address this, we implemented a strategy that reduces the overall memory footprint. This is based on the observation that all tasks along a column of the processor grid (cf. Fig.~\ref{fig:taskMpiComm}) require the same subgroup of $j$-indexed KS wavefunctions. Leveraging this, we implement a round-robin algorithm to transfer the j-indexed KS wavefunction data among the tasks in a column, as shown in Fig. \ref{fig:memMpiComm}. To elaborate, $i$-$j$ pairs are assigned to a task based on its position in the grid. Once the task completes the necessary calculations corresponding to the $i$-$j$ pairs for a given $j$, the wavefunction $j$ temporarily assigned to this task is sent to the next task in the column using the round-robin algorithm, reusing the pre-allocated memory.

While band parallelization is adopted for storing the KS wavefunctions in $\mathcal{S}-{\rm FE}$ basis, where the complete data for any wavefunction is local to a task, the parallelization employed for the $\mathcal{U}-{\rm FE}$ basis is based on domain decomposition, which is more ideal for the solution of the eigenvalue problem in the inner SCF iteration on electron density. To elaborate, in domain decomposition, the FE mesh is distributed across various MPI tasks, with each task holding the information for all KS wavefunctions in the finite-element cells local to the task. We note that DFT-FE framework, which handles all aspects of the computation in the inner SCF, is based on domain decomposition parallelization with good parallel scalability exhibited on both many-core CPU as well as hybrid CPU-GPU architectures~\cite{motamarri2020dft,das2022dft,das2019fast,GB2023}. Thus, the transfer of fields between $\mathcal{S}-{\rm FE}$ basis and $\mathcal{U}-{\rm FE}$ basis has to navigate the different parallelization strategies for $\mathcal{S}-{\rm FE}$ and $\mathcal{U}-{\rm FE}$ meshes. Importantly, the same set of processors are used to store the fields on both $\mathcal{S}-{\rm FE}$ and $\mathcal{U}-{\rm FE}$ basis. 

The main operation in the transfer of fields between $\mathcal{S}-{\rm FE}$ basis and $\mathcal{U}-{\rm FE}$ basis is an interpolation operation (cf. Eq.~\ref{eq:interpolate}). In order to transfer the KS wavefunctions from $\mathcal{U}-{\rm FE}$ basis to $\mathcal{S}-{\rm FE}$ basis, we first create MPI communication pattern that maps domain decomposition parallelization in $\mathcal{U}-{\rm FE}$ mesh to the band parallelization in $\mathcal{S}-{\rm FE}$ mesh. This entails mapping FE nodes of $\mathcal{S}-{\rm FE}$ to FE cells containing these spatial locations in the $\mathcal{U}-{\rm FE}$. This creates a communication pattern, where each task handling a set of KS wavefunctions in the $\mathcal{S}-{\rm FE}$ basis will gather the interpolated fields from the $\mathcal{U}-{\rm FE}$ mesh distributed over all tasks and collates the data. We note that the communication pattern only needs to be established at the beginning of calculations, and the interpolation and transfer of fields uses this communication pattern in every call of \texttt{updateExchange}. To improve the efficiency of the transfer operation, we also implemented overlapping compute and communication. While the communication to task $P$ of already computed interpolated values is happening, the computation of interpolated values to be sent to task $P+1$ can commence. We achieve this using asynchronous point-to-point communication calls ($\texttt{MPI}\_\texttt{Isend()}$, $\texttt{MPI}\_\texttt{Irecv()}$). A similar implementation is used for transferring fields from $\mathcal{S}-{\rm FE}$ basis to $\mathcal{U}-{\rm FE}$ basis. 
 
\section{Results and Discussion}
\label{ch:results}

We now present the accuracy, computational efficiency and parallel scalability of our implementation of the action of the exchange operator using Tucker tensor decomposition in the DFT-FE framework. We demonstrate the accuracy in ground-state energy on a range of small- to medium-sized benchmark materials systems involving both insulating and metallic systems. Further, we demonstrate the computational efficiency on Pt clusters of increasing sizes up till 1,584 electrons, and a large $\text{TiO}_{\text{2}}$ cluster with 3,648 electrons. We use ONCV SG 15 pseudo potentials \cite{PhysRevB.88.085117} and the PBE0 \cite{pbe0Citation} exchange-correlation functional for all these calculations. We employ Anderson mixing~\cite{anderson1965iterative} for the inner fixed-point iteration in electron density. We do not use mixing in the outer fixed-point iteration involving the Kohn-Sham orbitals in the construction of the exchange operator. In all the calculations, we use a Fermi-Dirac smearing corresponding to 500K.

\subsection{Accuracy benchmarks}
The ground state energy obtained from our implementation in DFT-FE is compared against Quantum Espresso (QE) \cite{giannozzi2009quantum} to validate the accuracy of the approach and implementation. We consider the hybrid functional DFT calculation in DFT-FE to be converged when the difference between the Fock exchange energy of successive outer SCF iterations is less than a pre-determined tolerance. We note that this is slightly different from the procedure followed by QE, in which the simulation runs until the difference in the total energy between successive outer SCFs is below a pre-determined tolerance. Hence, during the performance comparison, the simulation in both the codes is considered converged when the difference between the total energy between successive outer SCFs is below $10^{-5}$~Ha/atom. The discretization parameters in both QE and DFT-FE are set such that the discretization error is below $10^{-4}$~Ha/atom for the ground state energy obtained for PBE exchange-correlation functional. The parameters used to generate the $\mathcal{S}-{\rm FE}$ mesh are also set to be commensurate with discretization errors below $10^{-4}$~Ha/atom. The benchmark systems we use to demonstrate the accuracy include: i) Au-Pt dimer; ii) Benzamide molecule; iii) 19-atom Pt nanocluster (Pt-19); and iv) 38-atom Pt nanocluster (Pt-38). All QE studies are performed using a vacuum just sufficient for the error in ground-state energy to be below $10^{-4}$~Ha/atom. In DFT-FE large domains for non-periodic calculations are easily accessible by leveraging the spatial adaptivity in the finite-element basis. The computed ground-state energies for the various benchmark systems are reported in Table~\ref{tab:accuracy}. Notably, the ground-state energies from DFT-FE and QE agree to within the discretization errors of $10^{-4}$~Ha/atom. Also, the number of outer iterations for convergence, which involves updating the exchange operator, are similar for DFT-FE and QE. The specific details of the simulation are elaborated below. 

i) Au-Pt dimer: Au-Pt dimer with bond length of 4.753~Bohr in considered, which has 37~electrons. A plane-wave cutoff of 70~Ry is used for the KS wavefunctions in the QE simulation with a cell size of 40~Bohr. In the DFT-FE calculation, the $\mathcal{U}-{\rm FE}$ mesh is constructed using a mesh-size near atom ($msna$) of 0.4, a base mesh-size ($bms$) of 6.66, and  a finite-element polynomial order of 4. The  $\mathcal{S}-{\rm FE}$ mesh is constructed using a mesh-size near atom ($msna$) of 0.625, a base mesh-size ($bms$) of 10.0, and  a finite-element polynomial order of 4.

ii) Benzamide molecule: Benzamide molecule ($\text{C}_7\text{H}_7\text{NO}$) comprises of 46 electrons. A cutoff of 100 Ry is used for KS wavfunctions in QE, with a cell size of 90 Bohr. In the DFT-FE calculation, both $\mathcal{U}-FE$ and $\mathcal{S}-FE$ meshes are constructed using $bms$ of 10.0 and $msna$ of 1.2. The polynomial order for $\mathcal{U}-{\rm FE}$ is chosen to be 6, while the polynomial order for $\mathcal{S}-{\rm FE}$ is 5.   
 
iii) Pt -19 atom cluster: This system comprises of 342 electrons. The QE simulation is performed using a cutoff of 70 Ry for KS wavefunctions, with a cell size of 40 Bohr. The DFT-FE calculation is performed using a $\mathcal{U}-{\rm FE}$ with $bms$ 8.4, $msna$ 1.0, and polynomial order 6. The $\mathcal{S}-{\rm FE}$ is constructed with $bms$ 7.8, $msna$ 1.5, and polynomial order 6.

iv) Pt - 38 atom cluster: This system comprises of 684 electrons. The QE simulation is performed using a cutoff of 70 Ry for KS wavefunctions and a cell size of 50 Bohr. The DFT-FE calculation is performed using a $\mathcal{U}-{\rm FE}$ with $bms$ 7.5, $msna$ 1.0 and polynomial order 6. The $\mathcal{S}-{\rm FE}$ is constructed with $bms$ 11.0, $msna$ 1.5 and polynomial order 6.

\begin{table}[htbp!]
  \begin{center}
    \caption{Comparison of ground-state energies with PBE0 exchange-correlation functional for the various benchmark systems computed using DFT-FE and QE. $E_g$ denotes the ground-state energy in Ha. The difference in the computed ground-sate energies per atom using DFT-FE and QE is also tabulated.}
    \label{tab:accuracy}
    \begin{tabular}{|l|l|l|l|l|l|}
    \hline\hline
      System &
\multicolumn{2}{c|}{QE} &
\multicolumn{2}{c|}{DFT-FE} &
\multicolumn{1}{c||}{Diff in Energy} \\
\hline
 & $E_g$ (Ha) & \# outer & $E_g$ (Ha) &\# outer & (Ha/atom) \\ 
  &  & iterations &  & iterations &\\ 
\hline
Pt- Au dimer & -258.3823 & 10 &  -258.3826  & 6 &   1.89E-04 \\
\hline
Benzamide & -69.9217 & 8 & -69.9208 & 6 & 5.625E-05 \\
\hline
Pt - 19 & -2303.1141 & 14 & -2303.1144 & 11 & 1.38E-05 \\
\hline
Pt - 38 & -4606.269 & 27 & -4606.2655 & 22 & 9.22E-05 \\
 \hline\hline
    \end{tabular}
  \end{center}
\end{table}

\subsection{Performance}
The performance of our numerical implementation was measured on Stampede2's KNL nodes. Each Stampeded2 node has 96 GB of memory and 68 cores with four hardware threads. In our simulations, 24-36 MPI tasks per node were used to ensure sufficient memory is available per task. These simulations are performed using discretization parameters providing $\mathcal{O}(10^{-4})$ Ha/atom accuracy in ground-state energy. In order to demonstrate the computational efficiency, we choose three Pt nanoclusters of increasing sizes, containing 38 atoms (684 electrons), 63 atoms (1,134 electrons), and 88 atoms (1,584  electrons). Further, we consider a large $\text{TiO}_{\text{2}}$ cluster with 3,648 electrons.

\begin{figure}
     \centering
         \begin{subfigure}[b]{0.49\textwidth}
         \centering
         \includegraphics[width=\textwidth]{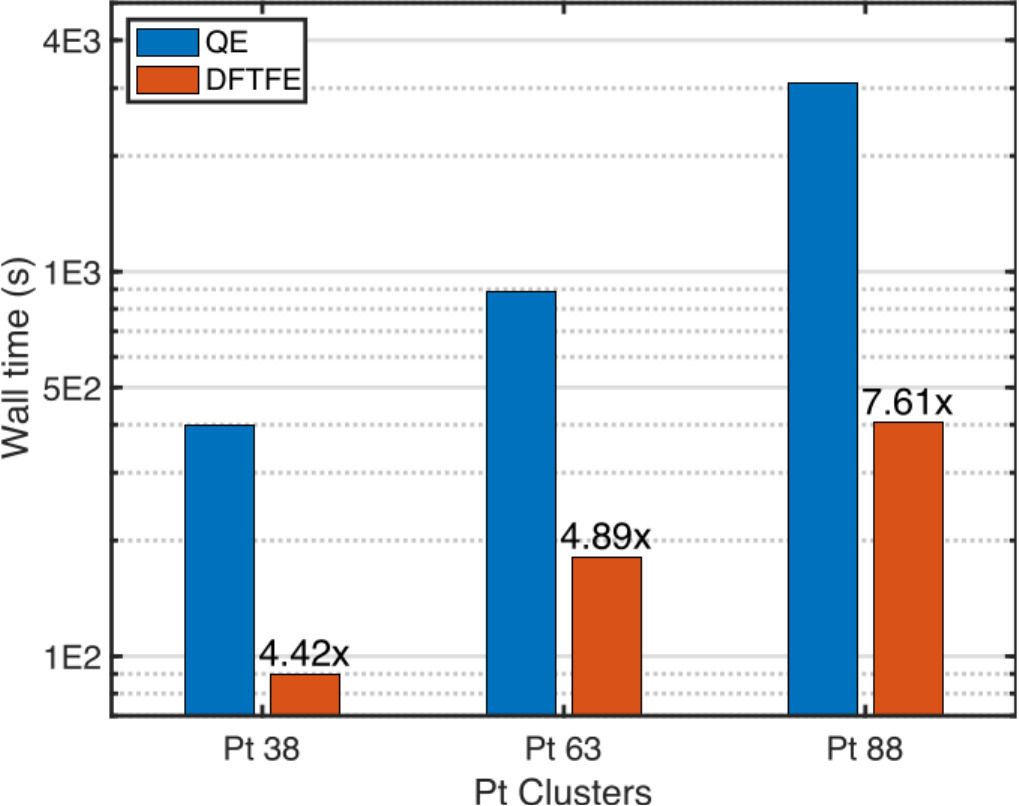}
         \caption{Wall time of \texttt{updateExchange} for different Pt clusters}
         \label{fig:exxPtCluster}
     \end{subfigure}
     \hfill
     \begin{subfigure}[b]{0.49\textwidth}
         \centering
         \includegraphics[width=\textwidth]{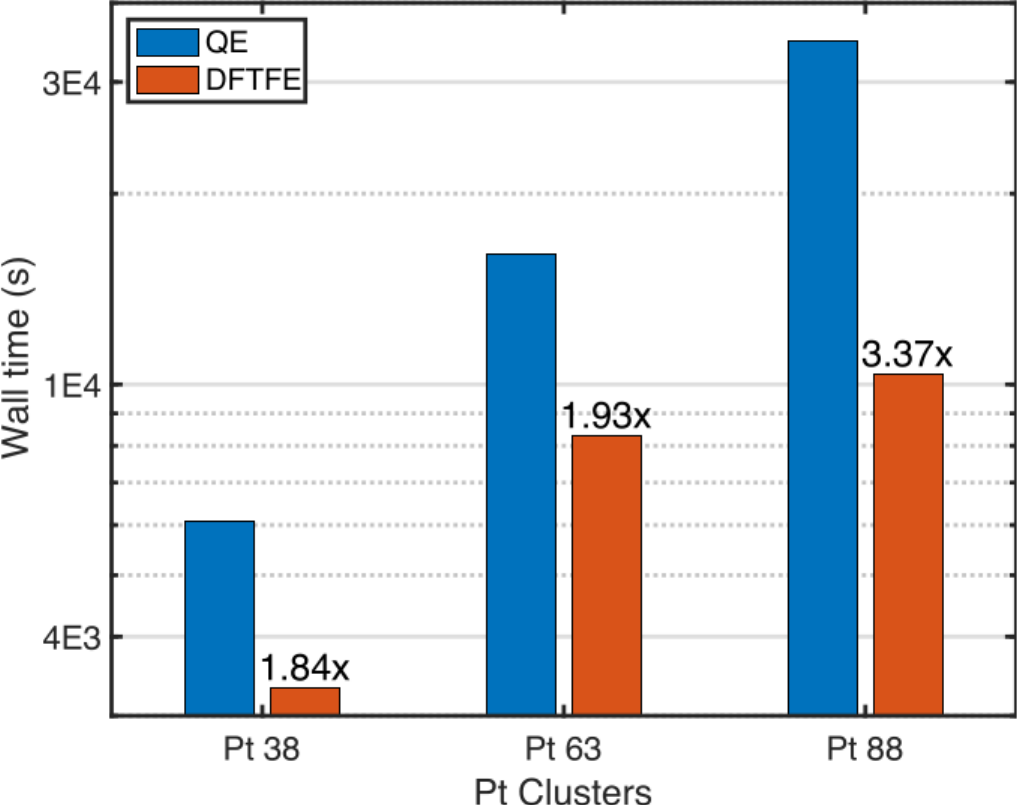}
         \caption{Wall time of ground state calculation for different Pt clusters}
         \label{fig:gsPtCluster}
     \end{subfigure}
     \caption{Wall time measurements for different Pt clusters}
        \label{fig:PtClusterTimings}
\end{figure}

We perform the full ground-state calculations for the benchmark Pt nanoclusters. The walltimes are measured using a tolerance of $10^{-5}$ Ha/atom in the convergence of outer SCF iteration for both QE and DFT-FE calculations. The performance of QE is heavily dependent on the band group parallelization. However, the extent of the band group parallelization was limited by the available memory. The comparison of walltimes for the construction of one ACE operator are shown in Fig. \ref{fig:exxPtCluster}. Figure~\ref{fig:gsPtCluster} shows the walltime comparison for the full ground-state calculations. We note that both DFT-FE and QE simulations are performed on the same number of nodes, thus walltimes are also representative of the relative computational efficiency. The Pt-38 simulations are run on 64 nodes, whereas the Pt-63 and Pt-88 simulations are run on 128 nodes. The speedup in DFT-FE relative to QE for the construction of ACE operator increases with increasing system size, reaching 7.6$\times$ for Pt-88. This translates to substantial improvement in the computational efficiency of the full ground-state calculation as well, resulting in $>$3$\times$ speedup for Pt-88.

\begin{figure}
     \centering
         \begin{subfigure}[b]{0.3\textwidth}
         \centering
         \includegraphics[width=\textwidth]{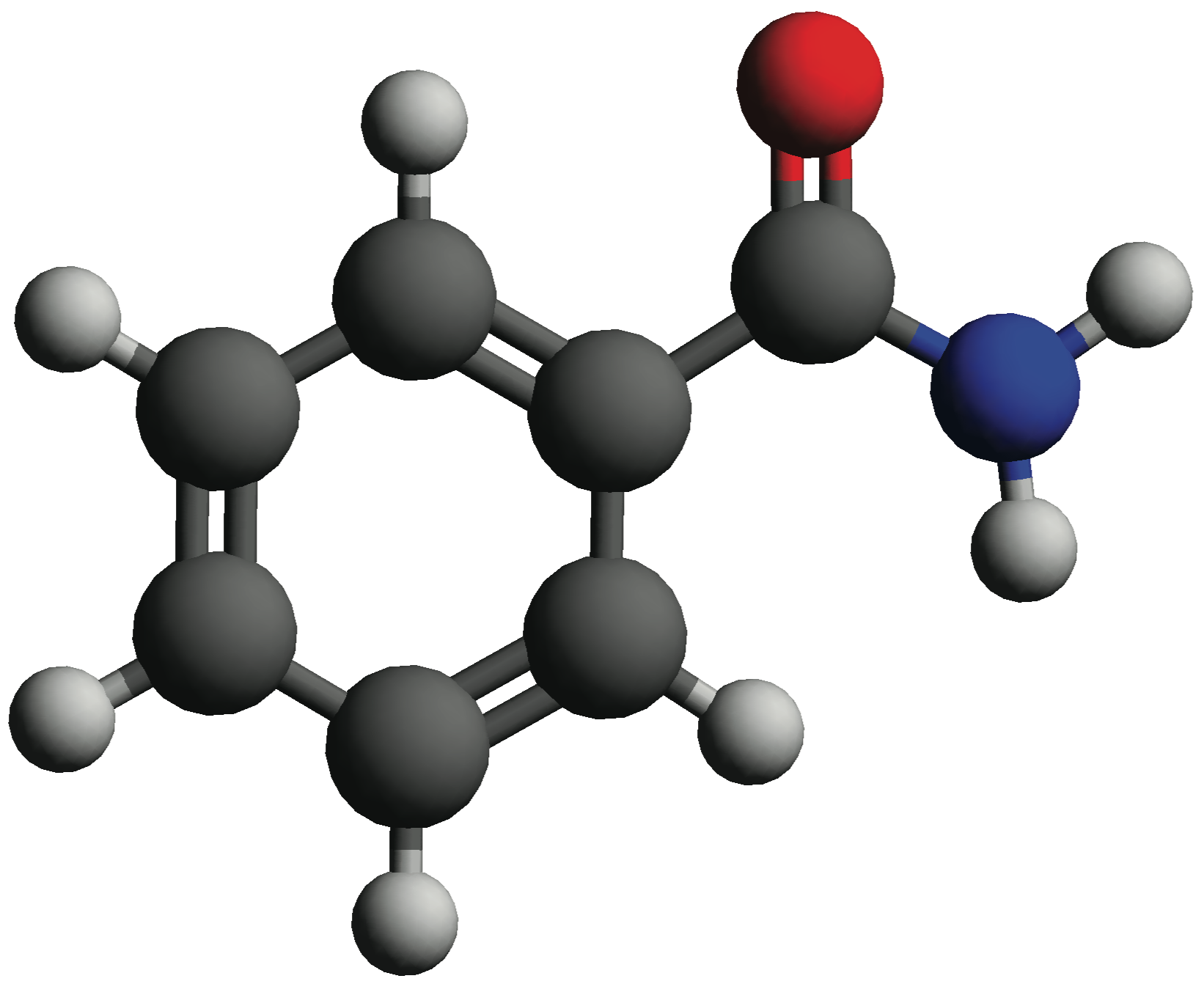}
         \caption{Schematic of Benzamide system}
         \label{fig:Benzamide}
     \end{subfigure}
     \hfill
     \begin{subfigure}[b]{0.3\textwidth}
         \centering
         \includegraphics[width=\textwidth]{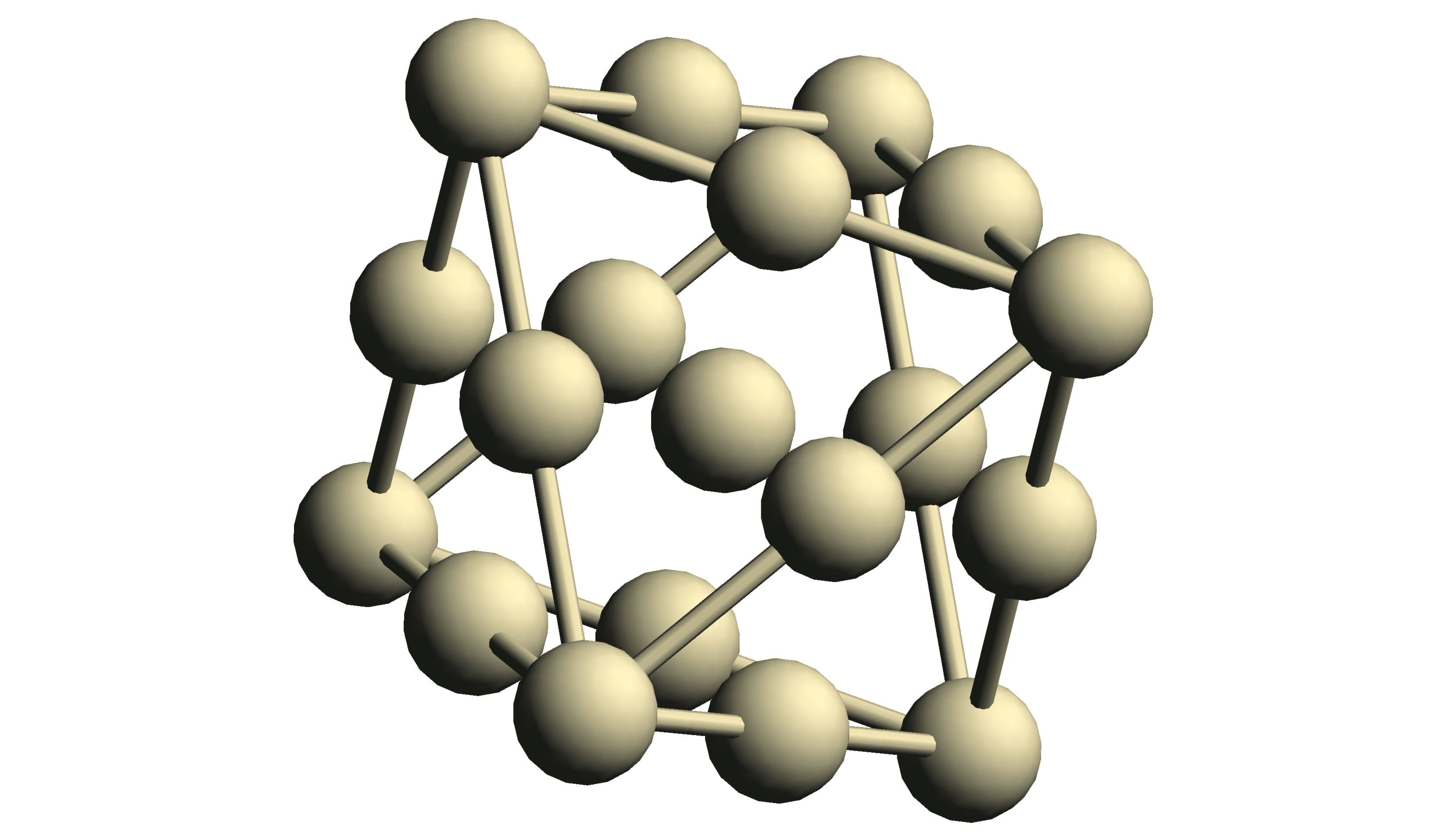}
         \caption{Schematic of Pt-19 atoms system; the bonds are shown for schematic purposed}
         \label{fig:Pt19}
     \end{subfigure}
     \hfill
     \begin{subfigure}[b]{0.3\textwidth}
         \centering
         \includegraphics[width=\textwidth]{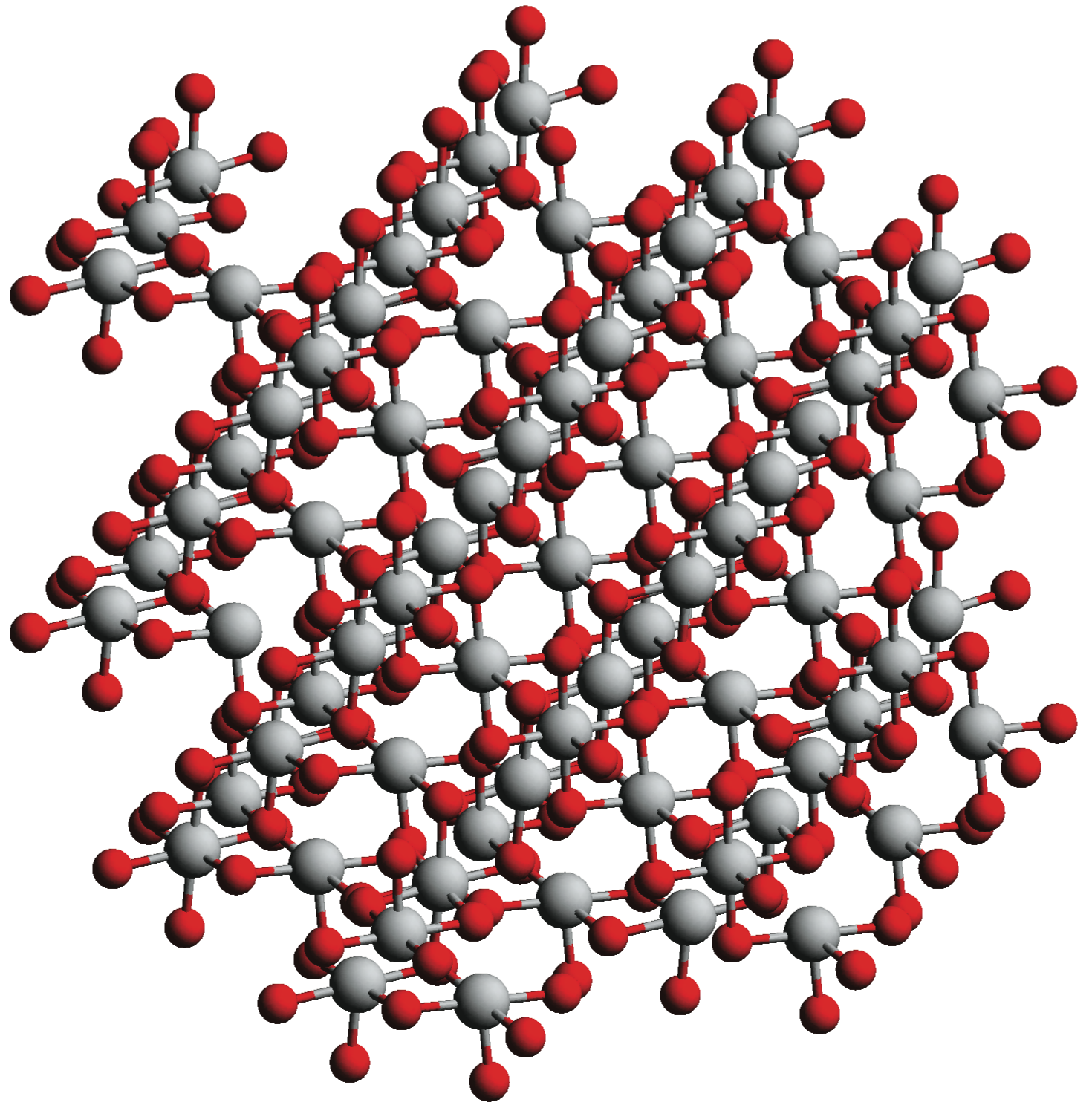}
         \caption{Schematic of $\text{TiO}_{\text{2}}$ cluster}
         \label{fig:tio2}
     \end{subfigure}
        \caption{Schematics of different benchmark systems used in the study}
        \label{fig:atom}
\end{figure}

The largest system for which the performance is measured is the $\text{TiO}_{\text{2}}$ system containing 480 atoms (depicted in Fig. \ref{fig:tio2}, 3,648 electrons and 1,900 wavefunctions). Both QE and DFT-FE calculations are performed on 192 nodes. As this is an expensive calculation, we only considered the comparison of the walltimes for the construction of the ACE operator. 
The results of the simulation are reported in Table \ref{tab:tio2Perf}, which shows a remarkable 11.3$\times$ speed-up over QE.

\begin{table}[htbp!]
  \begin{center}
    \caption{Comparison of walltimes for ACE construction for a 480 atom $\text{TiO}_{\text{2}}$ cluster.}
    \label{tab:tio2Perf}
    \begin{tabular}{||c c c c c||}
    \hline\hline
      System & No of Electrons & QE & DFT-FE & Speed-up  \\ 
       & & $(s)$ & $(s)$  &  \\ [0.5ex] 
      \hline\hline
$\text{TiO}_{\text{2}}$ & 3,648 & 21,390 & 1,892 & \textbf{11.3}  \\ [1ex] 
 \hline\hline
    \end{tabular}
  \end{center}
\end{table}

\subsection{Parallel scalability}
\subsubsection{Strong scaling}
\begin{figure}
     \centering
         \begin{subfigure}[b]{0.49\textwidth}
         \centering
         \includegraphics[width=\textwidth]{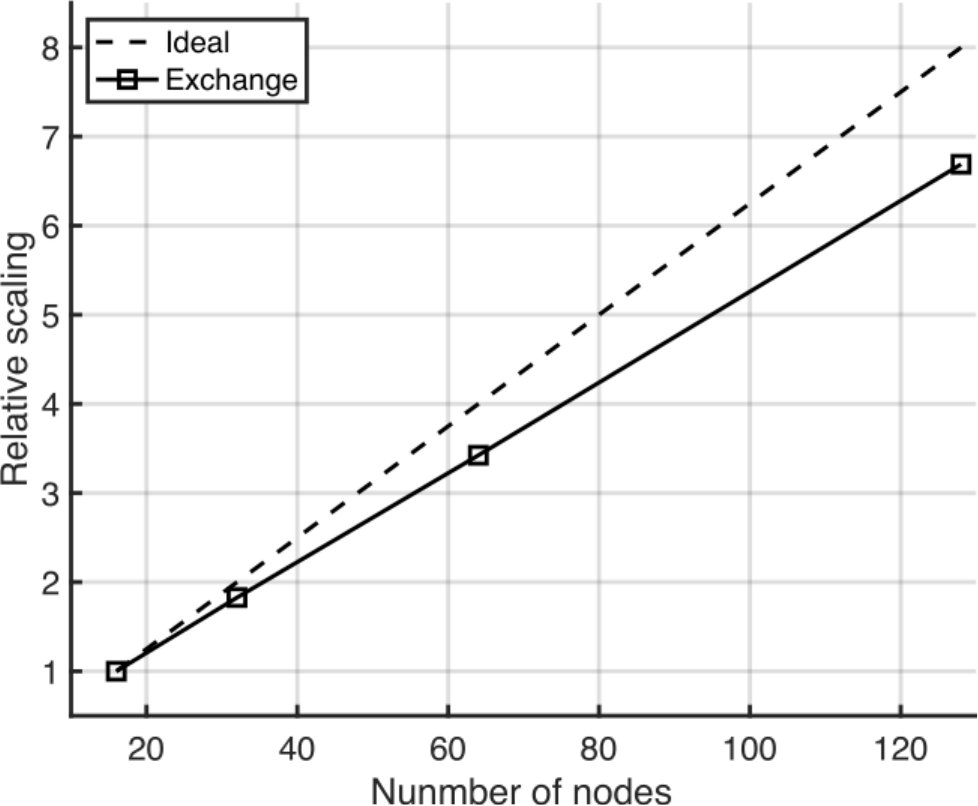}
         \caption{Strong scaling study of \texttt{updateExchange} }
         \label{fig:gsPt38}
     \end{subfigure}
     \hfill
     \begin{subfigure}[b]{0.49\textwidth}
         \centering
         \includegraphics[width=\textwidth]{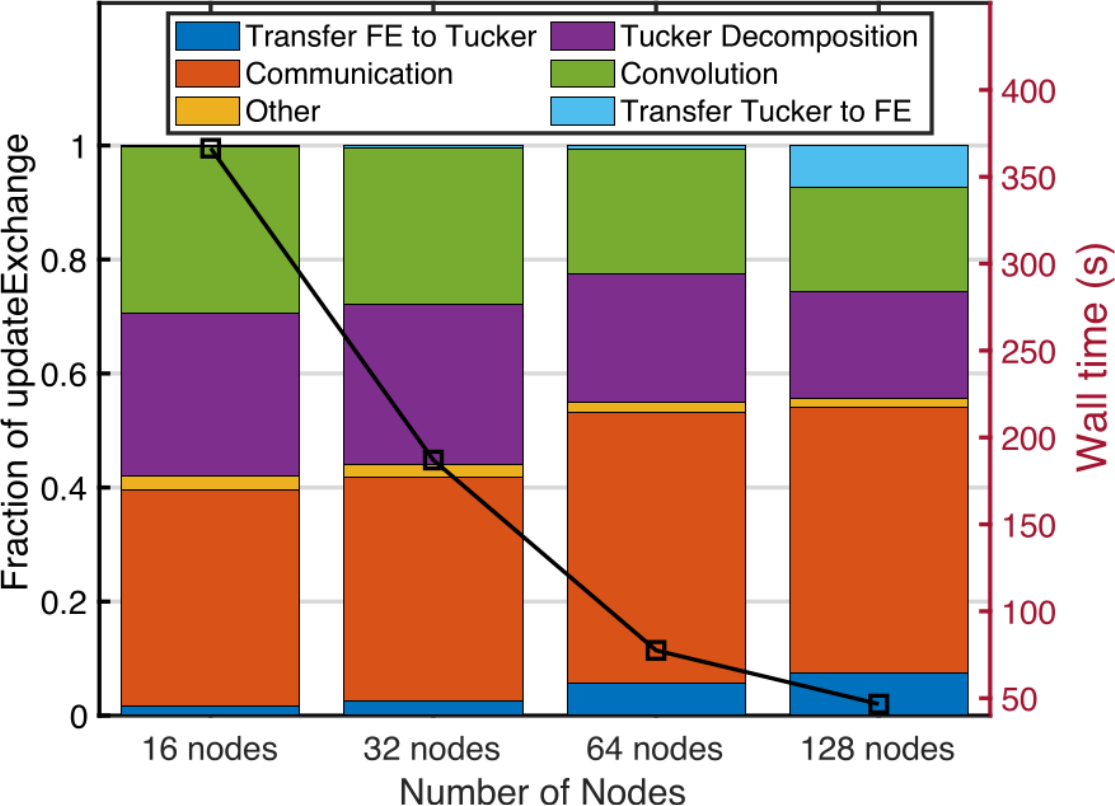}
         \caption{Performance breakdown for \texttt{updateExchange} for Pt 38 atom cluster on different number of nodes}
         \label{fig:perfBreak}
     \end{subfigure}
     \caption{Scaling analysis of \texttt{updateExchange} for Pt 38 atom cluster}
        \label{fig:PtClusterTimingsScaling}
\end{figure}

We use the Pt-38 nanocluster for demonstrating the parallel scalability of our implementation. The strong scaling is measured on nodes ranging from 16 nodes to 128 nodes. Figure~\ref{fig:gsPt38} shows the strong scaling of \texttt{updateExchange}. Notably, the walltime for this operation reduced from 380 seconds on 16 nodes to 56.8 seconds on 128 nodes, translating to a parallel efficiency of 84\%. We note that the number of finite-element nodes in $\mathcal{U}-{\rm FE}$ mesh in some processors is as low as 330 at 120 nodes, which demonstrates the excellent parallel scaling of the implementation. 

\subsubsection{Performance breakdown}
We also measured the time taken for various steps in the \texttt{updateExchange} operation. Since our implementation includes asynchronous compute by overlapping some aspects of compute and communication, the exact time taken by communication and computation can not be measured. Hence we measure the time taken by processor 0 during the computation steps and subtract it from the total time to estimate the time taken in communication that is not being overlapped. This data is presented in Fig.~\ref{fig:perfBreak}. We find that communication, Tucker decomposition, and convolution constitute the major fraction of walltime, and this is consistent across various number of nodes. Notably, the fraction of walltime for these steps does not change significantly as the number of nodes are increased, also demonstrating efficient parallel scaling across the various steps.

\section{Conclusion}\label{ch:conclusion}	
In this work, we employ a Tucker-tensor based approach to accelerate the evaluation of Fock exchange in a real-space finite-element basis, which is a systematically improvable and complete basis. The use of Tucker-tensor decomposition techniques allows the transformation of 3-D convolution integrals in the Fock exchange into a tensor product of 1-D convolution integrals, which results in a significant reduction in computational cost. This is achieved through a low-rank Tucker-tensor decomposition of the $\psi_i \psi_j$ products and a canonical decomposition of the $\frac{1}{r}$ kernel appearing in the Fock exchange. In addition, we have implemented a task/memory parallelization strategy over the $\psi_i \psi_j$ pairs that balance the peak memory and communication costs, in conjunction with strategies of overlapping compute-and-communication to reduce the communication cost and maintain good parallel scalability. While the proposed approach is compatible with any tensor structured basis, we have implemented the approach in DFT-FE---a massively parallel real-space DFT code based on adaptive higher order finite-element discretization---extending the capability of DFT-FE to treat hybrid XC functionals.

The accuracy and efficiency of the implementation have been demonstrated by comparing it against Quantum Espresso. The ground-state energies between DFT-FE and Quantum Espresso agree to $\sim10^{-4}$ Ha/atom for hybrid XC functional calculations using PBE0 on various benchmark systems. In terms of computational efficiency, the action of exchange operator on single-particle wavefunctions demonstrates a speedup of up to 11$\times$ in comparison to Quantum Espresso for a $\text{TiO}_{\text{2}}$ cluster comprising of 3,684 electrons. Further, the implementation exhibits good parallel scalability, even in the extreme parallel scaling regime. In the future, we aim to extend the present implementation to take advantage of GPU acceleration, alongside extending the formulation to periodic systems.    
\section*{Acknowledgments}
We gratefully acknowledge the support of Toyota Research Institute under the auspices of which this work was performed. This work used Stampede2 from the ACCESS program, which is supported by National Science Foundation grants \#2138259, \#2138286, \#2138307, \#2137603, and \#2138296.  

\bibliography{references}

\providecommand{\latin}[1]{#1}
\makeatletter
\providecommand{\doi}
  {\begingroup\let\do\@makeother\dospecials
  \catcode`\{=1 \catcode`\}=2 \doi@aux}
\providecommand{\doi@aux}[1]{\endgroup\texttt{#1}}
\makeatother
\providecommand*\mcitethebibliography{\thebibliography}
\csname @ifundefined\endcsname{endmcitethebibliography}  {\let\endmcitethebibliography\endthebibliography}{}
\begin{mcitethebibliography}{71}
\providecommand*\natexlab[1]{#1}
\providecommand*\mciteSetBstSublistMode[1]{}
\providecommand*\mciteSetBstMaxWidthForm[2]{}
\providecommand*\mciteBstWouldAddEndPuncttrue
  {\def\EndOfBibitem{\unskip.}}
\providecommand*\mciteBstWouldAddEndPunctfalse
  {\let\EndOfBibitem\relax}
\providecommand*\mciteSetBstMidEndSepPunct[3]{}
\providecommand*\mciteSetBstSublistLabelBeginEnd[3]{}
\providecommand*\EndOfBibitem{}
\mciteSetBstSublistMode{f}
\mciteSetBstMaxWidthForm{subitem}{(\alph{mcitesubitemcount})}
\mciteSetBstSublistLabelBeginEnd
  {\mcitemaxwidthsubitemform\space}
  {\relax}
  {\relax}

\bibitem[Hohenberg and Kohn(1964)Hohenberg, and Kohn]{PhysRev.136.B864}
Hohenberg,~P.; Kohn,~W. Inhomogeneous Electron Gas. \emph{Phys. Rev.} \textbf{1964}, \emph{136}, B864--B871\relax
\mciteBstWouldAddEndPuncttrue
\mciteSetBstMidEndSepPunct{\mcitedefaultmidpunct}
{\mcitedefaultendpunct}{\mcitedefaultseppunct}\relax
\EndOfBibitem
\bibitem[Kohn and Sham(1965)Kohn, and Sham]{PhysRev.140.A1133}
Kohn,~W.; Sham,~L.~J. Self-Consistent Equations Including Exchange and Correlation Effects. \emph{Phys. Rev.} \textbf{1965}, \emph{140}, A1133--A1138\relax
\mciteBstWouldAddEndPuncttrue
\mciteSetBstMidEndSepPunct{\mcitedefaultmidpunct}
{\mcitedefaultendpunct}{\mcitedefaultseppunct}\relax
\EndOfBibitem
\bibitem[Martin(2020)]{martin2020electronic}
Martin,~R.~M. \emph{Electronic structure: basic theory and practical methods}; Cambridge university press, 2020\relax
\mciteBstWouldAddEndPuncttrue
\mciteSetBstMidEndSepPunct{\mcitedefaultmidpunct}
{\mcitedefaultendpunct}{\mcitedefaultseppunct}\relax
\EndOfBibitem
\bibitem[Hafner \latin{et~al.}(2006)Hafner, Wolverton, and Ceder]{hafner2006toward}
Hafner,~J.; Wolverton,~C.; Ceder,~G. Toward computational materials design: the impact of density functional theory on materials research. \emph{MRS bulletin} \textbf{2006}, \emph{31}, 659--668\relax
\mciteBstWouldAddEndPuncttrue
\mciteSetBstMidEndSepPunct{\mcitedefaultmidpunct}
{\mcitedefaultendpunct}{\mcitedefaultseppunct}\relax
\EndOfBibitem
\bibitem[Becke(2014)]{becke2014perspective}
Becke,~A.~D. Perspective: Fifty years of density-functional theory in chemical physics. \emph{The Journal of chemical physics} \textbf{2014}, \emph{140}, 18A301\relax
\mciteBstWouldAddEndPuncttrue
\mciteSetBstMidEndSepPunct{\mcitedefaultmidpunct}
{\mcitedefaultendpunct}{\mcitedefaultseppunct}\relax
\EndOfBibitem
\bibitem[Mardirossian and Head-Gordon(2017)Mardirossian, and Head-Gordon]{mardirossian2017thirty}
Mardirossian,~N.; Head-Gordon,~M. Thirty years of density functional theory in computational chemistry: an overview and extensive assessment of 200 density functionals. \emph{Molecular Physics} \textbf{2017}, \emph{115}, 2315--2372\relax
\mciteBstWouldAddEndPuncttrue
\mciteSetBstMidEndSepPunct{\mcitedefaultmidpunct}
{\mcitedefaultendpunct}{\mcitedefaultseppunct}\relax
\EndOfBibitem
\bibitem[Kim \latin{et~al.}(2006)Kim, Lee, Kang, Ko, Yum, Fantacci, De~Angelis, Di~Censo, Nazeeruddin, and Gr{\"a}tzel]{kim2006molecular}
Kim,~S.; Lee,~J.~K.; Kang,~S.~O.; Ko,~J.; Yum,~J.-H.; Fantacci,~S.; De~Angelis,~F.; Di~Censo,~D.; Nazeeruddin,~M.~K.; Gr{\"a}tzel,~M. Molecular engineering of organic sensitizers for solar cell applications. \emph{Journal of the American Chemical Society} \textbf{2006}, \emph{128}, 16701--16707\relax
\mciteBstWouldAddEndPuncttrue
\mciteSetBstMidEndSepPunct{\mcitedefaultmidpunct}
{\mcitedefaultendpunct}{\mcitedefaultseppunct}\relax
\EndOfBibitem
\bibitem[Jhi \latin{et~al.}(2000)Jhi, Louie, and Cohen]{jhi2000electronic}
Jhi,~S.-H.; Louie,~S.~G.; Cohen,~M.~L. Electronic properties of oxidized carbon nanotubes. \emph{Physical Review Letters} \textbf{2000}, \emph{85}, 1710\relax
\mciteBstWouldAddEndPuncttrue
\mciteSetBstMidEndSepPunct{\mcitedefaultmidpunct}
{\mcitedefaultendpunct}{\mcitedefaultseppunct}\relax
\EndOfBibitem
\bibitem[Kohn and Sham(1965)Kohn, and Sham]{kohn1965self}
Kohn,~W.; Sham,~L.~J. Self-consistent equations including exchange and correlation effects. \emph{Physical review} \textbf{1965}, \emph{140}, A1133\relax
\mciteBstWouldAddEndPuncttrue
\mciteSetBstMidEndSepPunct{\mcitedefaultmidpunct}
{\mcitedefaultendpunct}{\mcitedefaultseppunct}\relax
\EndOfBibitem
\bibitem[Perdew \latin{et~al.}(1996)Perdew, Burke, and Ernzerhof]{PhysRevLett.77.3865}
Perdew,~J.~P.; Burke,~K.; Ernzerhof,~M. Generalized Gradient Approximation Made Simple. \emph{Phys. Rev. Lett.} \textbf{1996}, \emph{77}, 3865--3868\relax
\mciteBstWouldAddEndPuncttrue
\mciteSetBstMidEndSepPunct{\mcitedefaultmidpunct}
{\mcitedefaultendpunct}{\mcitedefaultseppunct}\relax
\EndOfBibitem
\bibitem[Becke(1993)]{becke98density}
Becke,~A.~D. {Density‐functional thermochemistry. III. The role of exact exchange}. \emph{The Journal of Chemical Physics} \textbf{1993}, \emph{98}, 5648--5652\relax
\mciteBstWouldAddEndPuncttrue
\mciteSetBstMidEndSepPunct{\mcitedefaultmidpunct}
{\mcitedefaultendpunct}{\mcitedefaultseppunct}\relax
\EndOfBibitem
\bibitem[Perdew \latin{et~al.}(1996)Perdew, Ernzerhof, and Burke]{pbe0Citation}
Perdew,~J.~P.; Ernzerhof,~M.; Burke,~K. {Rationale for mixing exact exchange with density functional approximations}. \emph{The Journal of Chemical Physics} \textbf{1996}, \emph{105}, 9982--9985\relax
\mciteBstWouldAddEndPuncttrue
\mciteSetBstMidEndSepPunct{\mcitedefaultmidpunct}
{\mcitedefaultendpunct}{\mcitedefaultseppunct}\relax
\EndOfBibitem
\bibitem[Heyd \latin{et~al.}(2003)Heyd, Scuseria, and Ernzerhof]{doi:10.1063/1.1564060}
Heyd,~J.; Scuseria,~G.~E.; Ernzerhof,~M. Hybrid functionals based on a screened Coulomb potential. \emph{The Journal of Chemical Physics} \textbf{2003}, \emph{118}, 8207--8215\relax
\mciteBstWouldAddEndPuncttrue
\mciteSetBstMidEndSepPunct{\mcitedefaultmidpunct}
{\mcitedefaultendpunct}{\mcitedefaultseppunct}\relax
\EndOfBibitem
\bibitem[Krukau \latin{et~al.}(2006)Krukau, Vydrov, Izmaylov, and Scuseria]{krukau2006influence}
Krukau,~A.~V.; Vydrov,~O.~A.; Izmaylov,~A.~F.; Scuseria,~G.~E. Influence of the exchange screening parameter on the performance of screened hybrid functionals. \emph{The Journal of chemical physics} \textbf{2006}, \emph{125}\relax
\mciteBstWouldAddEndPuncttrue
\mciteSetBstMidEndSepPunct{\mcitedefaultmidpunct}
{\mcitedefaultendpunct}{\mcitedefaultseppunct}\relax
\EndOfBibitem
\bibitem[Finazzi \latin{et~al.}(2008)Finazzi, Di~Valentin, Pacchioni, and Selloni]{finazzi2008excess}
Finazzi,~E.; Di~Valentin,~C.; Pacchioni,~G.; Selloni,~A. {Excess electron states in reduced bulk anatase $TiO_2$: comparison of standard GGA, GGA+ U, and hybrid DFT calculations}. \emph{The Journal of chemical physics} \textbf{2008}, \emph{129}, 154113\relax
\mciteBstWouldAddEndPuncttrue
\mciteSetBstMidEndSepPunct{\mcitedefaultmidpunct}
{\mcitedefaultendpunct}{\mcitedefaultseppunct}\relax
\EndOfBibitem
\bibitem[Da~Silva \latin{et~al.}(2007)Da~Silva, Ganduglia-Pirovano, Sauer, Bayer, and Kresse]{da2007hybrid}
Da~Silva,~J.~L.; Ganduglia-Pirovano,~M.~V.; Sauer,~J.; Bayer,~V.; Kresse,~G. Hybrid functionals applied to rare-earth oxides: The example of ceria. \emph{Physical Review B} \textbf{2007}, \emph{75}, 045121\relax
\mciteBstWouldAddEndPuncttrue
\mciteSetBstMidEndSepPunct{\mcitedefaultmidpunct}
{\mcitedefaultendpunct}{\mcitedefaultseppunct}\relax
\EndOfBibitem
\bibitem[Garza and Scuseria(2016)Garza, and Scuseria]{garza2016predicting}
Garza,~A.~J.; Scuseria,~G.~E. Predicting band gaps with hybrid density functionals. \emph{The journal of physical chemistry letters} \textbf{2016}, \emph{7}, 4165--4170\relax
\mciteBstWouldAddEndPuncttrue
\mciteSetBstMidEndSepPunct{\mcitedefaultmidpunct}
{\mcitedefaultendpunct}{\mcitedefaultseppunct}\relax
\EndOfBibitem
\bibitem[Heyd \latin{et~al.}(2005)Heyd, Peralta, Scuseria, and Martin]{heyd2005energy}
Heyd,~J.; Peralta,~J.~E.; Scuseria,~G.~E.; Martin,~R.~L. Energy band gaps and lattice parameters evaluated with the Heyd-Scuseria-Ernzerhof screened hybrid functional. \emph{The Journal of chemical physics} \textbf{2005}, \emph{123}, 174101\relax
\mciteBstWouldAddEndPuncttrue
\mciteSetBstMidEndSepPunct{\mcitedefaultmidpunct}
{\mcitedefaultendpunct}{\mcitedefaultseppunct}\relax
\EndOfBibitem
\bibitem[Muscat \latin{et~al.}(2001)Muscat, Wander, and Harrison]{muscat2001prediction}
Muscat,~J.; Wander,~A.; Harrison,~N. On the prediction of band gaps from hybrid functional theory. \emph{Chemical Physics Letters} \textbf{2001}, \emph{342}, 397--401\relax
\mciteBstWouldAddEndPuncttrue
\mciteSetBstMidEndSepPunct{\mcitedefaultmidpunct}
{\mcitedefaultendpunct}{\mcitedefaultseppunct}\relax
\EndOfBibitem
\bibitem[Salzner \latin{et~al.}(1997)Salzner, Lagowski, Pickup, and Poirier]{salzner1997design}
Salzner,~U.; Lagowski,~J.; Pickup,~P.; Poirier,~R. Design of low band gap polymers employing density functional theory—hybrid functionals ameliorate band gap problem. \emph{Journal of computational chemistry} \textbf{1997}, \emph{18}, 1943--1953\relax
\mciteBstWouldAddEndPuncttrue
\mciteSetBstMidEndSepPunct{\mcitedefaultmidpunct}
{\mcitedefaultendpunct}{\mcitedefaultseppunct}\relax
\EndOfBibitem
\bibitem[Xiao \latin{et~al.}(2011)Xiao, Tahir-Kheli, and Goddard~III]{xiao2011accurate}
Xiao,~H.; Tahir-Kheli,~J.; Goddard~III,~W.~A. Accurate band gaps for semiconductors from density functional theory. \emph{The Journal of Physical Chemistry Letters} \textbf{2011}, \emph{2}, 212--217\relax
\mciteBstWouldAddEndPuncttrue
\mciteSetBstMidEndSepPunct{\mcitedefaultmidpunct}
{\mcitedefaultendpunct}{\mcitedefaultseppunct}\relax
\EndOfBibitem
\bibitem[Seidl \latin{et~al.}(1996)Seidl, G{\"o}rling, Vogl, Majewski, and Levy]{seidl1996generalized}
Seidl,~A.; G{\"o}rling,~A.; Vogl,~P.; Majewski,~J.~A.; Levy,~M. Generalized Kohn-Sham schemes and the band-gap problem. \emph{Physical Review B} \textbf{1996}, \emph{53}, 3764\relax
\mciteBstWouldAddEndPuncttrue
\mciteSetBstMidEndSepPunct{\mcitedefaultmidpunct}
{\mcitedefaultendpunct}{\mcitedefaultseppunct}\relax
\EndOfBibitem
\bibitem[Lin(2016)]{lin2016adaptively}
Lin,~L. Adaptively compressed exchange operator. \emph{Journal of chemical theory and computation} \textbf{2016}, \emph{12}, 2242--2249\relax
\mciteBstWouldAddEndPuncttrue
\mciteSetBstMidEndSepPunct{\mcitedefaultmidpunct}
{\mcitedefaultendpunct}{\mcitedefaultseppunct}\relax
\EndOfBibitem
\bibitem[Hu \latin{et~al.}(2020)Hu, Liu, Li, Ding, Yang, and Yang]{hu2020accelerating}
Hu,~W.; Liu,~J.; Li,~Y.; Ding,~Z.; Yang,~C.; Yang,~J. Accelerating excitation energy computation in molecules and solids within linear-response time-dependent density functional theory via interpolative separable density fitting decomposition. \emph{Journal of Chemical Theory and Computation} \textbf{2020}, \emph{16}, 964--973\relax
\mciteBstWouldAddEndPuncttrue
\mciteSetBstMidEndSepPunct{\mcitedefaultmidpunct}
{\mcitedefaultendpunct}{\mcitedefaultseppunct}\relax
\EndOfBibitem
\bibitem[Hu \latin{et~al.}(2017)Hu, Lin, and Yang]{doi:10.1021/acs.jctc.7b00807}
Hu,~W.; Lin,~L.; Yang,~C. Interpolative Separable Density Fitting Decomposition for Accelerating Hybrid Density Functional Calculations with Applications to Defects in Silicon. \emph{Journal of Chemical Theory and Computation} \textbf{2017}, \emph{13}, 5420--5431, PMID: 28960982\relax
\mciteBstWouldAddEndPuncttrue
\mciteSetBstMidEndSepPunct{\mcitedefaultmidpunct}
{\mcitedefaultendpunct}{\mcitedefaultseppunct}\relax
\EndOfBibitem
\bibitem[Ko \latin{et~al.}(2020)Ko, Jia, Santra, Wu, Car, and DiStasio~Jr]{ko2020enabling}
Ko,~H.-Y.; Jia,~J.; Santra,~B.; Wu,~X.; Car,~R.; DiStasio~Jr,~R.~A. Enabling large-scale condensed-phase hybrid density functional theory based ab initio molecular dynamics. 1. theory, algorithm, and performance. \emph{Journal of Chemical Theory and Computation} \textbf{2020}, \emph{16}, 3757--3785\relax
\mciteBstWouldAddEndPuncttrue
\mciteSetBstMidEndSepPunct{\mcitedefaultmidpunct}
{\mcitedefaultendpunct}{\mcitedefaultseppunct}\relax
\EndOfBibitem
\bibitem[Ko \latin{et~al.}(2021)Ko, Santra, and DiStasio~Jr]{ko2021enabling}
Ko,~H.-Y.; Santra,~B.; DiStasio~Jr,~R.~A. Enabling Large-Scale Condensed-Phase Hybrid Density Functional Theory-Based Ab Initio Molecular Dynamics II: Extensions to the Isobaric--Isoenthalpic and Isobaric--Isothermal Ensembles. \emph{Journal of Chemical Theory and Computation} \textbf{2021}, \emph{17}, 7789--7813\relax
\mciteBstWouldAddEndPuncttrue
\mciteSetBstMidEndSepPunct{\mcitedefaultmidpunct}
{\mcitedefaultendpunct}{\mcitedefaultseppunct}\relax
\EndOfBibitem
\bibitem[Rettig \latin{et~al.}()Rettig, Lee, and Head-Gordon]{rettig2023even}
Rettig,~A.; Lee,~J.; Head-Gordon,~M. Even Faster Exact Exchange for Solids via Tensor Hypercontraction. \emph{Journal of Chemical Theory and Computation} PMID: 37586065\relax
\mciteBstWouldAddEndPuncttrue
\mciteSetBstMidEndSepPunct{\mcitedefaultmidpunct}
{\mcitedefaultendpunct}{\mcitedefaultseppunct}\relax
\EndOfBibitem
\bibitem[Apra \latin{et~al.}(2020)Apra, Bylaska, De~Jong, Govind, Kowalski, Straatsma, Valiev, van Dam, Alexeev, Anchell, \latin{et~al.} others]{apra2020nwchem}
Apra,~E.; Bylaska,~E.~J.; De~Jong,~W.~A.; Govind,~N.; Kowalski,~K.; Straatsma,~T.~P.; Valiev,~M.; van Dam,~H.~J.; Alexeev,~Y.; Anchell,~J.; others NWChem: Past, present, and future. \emph{The Journal of chemical physics} \textbf{2020}, \emph{152}, 184102\relax
\mciteBstWouldAddEndPuncttrue
\mciteSetBstMidEndSepPunct{\mcitedefaultmidpunct}
{\mcitedefaultendpunct}{\mcitedefaultseppunct}\relax
\EndOfBibitem
\bibitem[Blum \latin{et~al.}(2009)Blum, Gehrke, Hanke, Havu, Havu, Ren, Reuter, and Scheffler]{blum2009ab}
Blum,~V.; Gehrke,~R.; Hanke,~F.; Havu,~P.; Havu,~V.; Ren,~X.; Reuter,~K.; Scheffler,~M. Ab initio molecular simulations with numeric atom-centered orbitals. \emph{Computer Physics Communications} \textbf{2009}, \emph{180}, 2175--2196\relax
\mciteBstWouldAddEndPuncttrue
\mciteSetBstMidEndSepPunct{\mcitedefaultmidpunct}
{\mcitedefaultendpunct}{\mcitedefaultseppunct}\relax
\EndOfBibitem
\bibitem[Frisch(2009)]{frisch2009gaussian}
Frisch,~A. gaussian 09W Reference. \emph{Wallingford, USA, 25p} \textbf{2009}, \relax
\mciteBstWouldAddEndPunctfalse
\mciteSetBstMidEndSepPunct{\mcitedefaultmidpunct}
{}{\mcitedefaultseppunct}\relax
\EndOfBibitem
\bibitem[Giannozzi \latin{et~al.}(2009)Giannozzi, Baroni, Bonini, Calandra, Car, Cavazzoni, Ceresoli, Chiarotti, Cococcioni, Dabo, \latin{et~al.} others]{giannozzi2009quantum}
Giannozzi,~P.; Baroni,~S.; Bonini,~N.; Calandra,~M.; Car,~R.; Cavazzoni,~C.; Ceresoli,~D.; Chiarotti,~G.~L.; Cococcioni,~M.; Dabo,~I.; others QUANTUM ESPRESSO: a modular and open-source software project for quantum simulations of materials. \emph{Journal of physics: Condensed matter} \textbf{2009}, \emph{21}, 395502\relax
\mciteBstWouldAddEndPuncttrue
\mciteSetBstMidEndSepPunct{\mcitedefaultmidpunct}
{\mcitedefaultendpunct}{\mcitedefaultseppunct}\relax
\EndOfBibitem
\bibitem[Gonze \latin{et~al.}(2009)Gonze, Amadon, Anglade, Beuken, Bottin, Boulanger, Bruneval, Caliste, Caracas, Côté, Deutsch, Genovese, Ghosez, Giantomassi, Goedecker, Hamann, Hermet, Jollet, Jomard, Leroux, Mancini, Mazevet, Oliveira, Onida, Pouillon, Rangel, Rignanese, Sangalli, Shaltaf, Torrent, Verstraete, Zerah, and Zwanziger]{Gonze20092582}
Gonze,~X.; Amadon,~B.; Anglade,~P.-M.; Beuken,~J.-M.; Bottin,~F.; Boulanger,~P.; Bruneval,~F.; Caliste,~D.; Caracas,~R.; Côté,~M.; Deutsch,~T.; Genovese,~L.; Ghosez,~P.; Giantomassi,~M.; Goedecker,~S.; Hamann,~D.; Hermet,~P.; Jollet,~F.; Jomard,~G.; Leroux,~S.; Mancini,~M.; Mazevet,~S.; Oliveira,~M.; Onida,~G.; Pouillon,~Y.; Rangel,~T.; Rignanese,~G.-M.; Sangalli,~D.; Shaltaf,~R.; Torrent,~M.; Verstraete,~M.; Zerah,~G.; Zwanziger,~J. ABINIT: First-principles approach to material and nanosystem properties. \emph{Computer Physics Communications} \textbf{2009}, \emph{180}, 2582--2615, cited By 1869\relax
\mciteBstWouldAddEndPuncttrue
\mciteSetBstMidEndSepPunct{\mcitedefaultmidpunct}
{\mcitedefaultendpunct}{\mcitedefaultseppunct}\relax
\EndOfBibitem
\bibitem[Hafner(2008)]{hafner2008ab}
Hafner,~J. Ab-initio simulations of materials using VASP: Density-functional theory and beyond. \emph{Journal of computational chemistry} \textbf{2008}, \emph{29}, 2044--2078\relax
\mciteBstWouldAddEndPuncttrue
\mciteSetBstMidEndSepPunct{\mcitedefaultmidpunct}
{\mcitedefaultendpunct}{\mcitedefaultseppunct}\relax
\EndOfBibitem
\bibitem[Motamarri \latin{et~al.}(2013)Motamarri, Nowak, Leiter, Knap, and Gavini]{motamarri2013higher}
Motamarri,~P.; Nowak,~M.~R.; Leiter,~K.; Knap,~J.; Gavini,~V. Higher-order adaptive finite-element methods for Kohn--Sham density functional theory. \emph{Journal of Computational Physics} \textbf{2013}, \emph{253}, 308--343\relax
\mciteBstWouldAddEndPuncttrue
\mciteSetBstMidEndSepPunct{\mcitedefaultmidpunct}
{\mcitedefaultendpunct}{\mcitedefaultseppunct}\relax
\EndOfBibitem
\bibitem[Pask and Sterne(2005)Pask, and Sterne]{pask2005finite}
Pask,~J.; Sterne,~P. Finite element methods in ab initio electronic structure calculations. \emph{Modelling and Simulation in Materials Science and Engineering} \textbf{2005}, \emph{13}, R71\relax
\mciteBstWouldAddEndPuncttrue
\mciteSetBstMidEndSepPunct{\mcitedefaultmidpunct}
{\mcitedefaultendpunct}{\mcitedefaultseppunct}\relax
\EndOfBibitem
\bibitem[Motamarri \latin{et~al.}(2020)Motamarri, Das, Rudraraju, Ghosh, Davydov, and Gavini]{motamarri2020dft}
Motamarri,~P.; Das,~S.; Rudraraju,~S.; Ghosh,~K.; Davydov,~D.; Gavini,~V. {DFT-FE--A massively parallel adaptive finite-element code for large-scale density functional theory calculations}. \emph{Computer Physics Communications} \textbf{2020}, \emph{246}, 106853\relax
\mciteBstWouldAddEndPuncttrue
\mciteSetBstMidEndSepPunct{\mcitedefaultmidpunct}
{\mcitedefaultendpunct}{\mcitedefaultseppunct}\relax
\EndOfBibitem
\bibitem[Das \latin{et~al.}(2022)Das, Motamarri, Subramanian, Rogers, and Gavini]{das2022dft}
Das,~S.; Motamarri,~P.; Subramanian,~V.; Rogers,~D.~M.; Gavini,~V. DFT-FE 1.0: A massively parallel hybrid CPU-GPU density functional theory code using finite-element discretization. \emph{Computer Physics Communications} \textbf{2022}, \emph{280}, 108473\relax
\mciteBstWouldAddEndPuncttrue
\mciteSetBstMidEndSepPunct{\mcitedefaultmidpunct}
{\mcitedefaultendpunct}{\mcitedefaultseppunct}\relax
\EndOfBibitem
\bibitem[Das \latin{et~al.}(2019)Das, Motamarri, Gavini, Turcksin, Li, and Leback]{das2019fast}
Das,~S.; Motamarri,~P.; Gavini,~V.; Turcksin,~B.; Li,~Y.~W.; Leback,~B. Fast, scalable and accurate finite-element based ab initio calculations using mixed precision computing: 46 PFLOPS simulation of a metallic dislocation system. In Proceedings of the International Conference for High Performance Computing, Networking, Storage and Analysis (SC'19). 2019; pp 1--11\relax
\mciteBstWouldAddEndPuncttrue
\mciteSetBstMidEndSepPunct{\mcitedefaultmidpunct}
{\mcitedefaultendpunct}{\mcitedefaultseppunct}\relax
\EndOfBibitem
\bibitem[Das \latin{et~al.}(2023)Das, Bikash, Subramanian, Panigrahi, Motamarri, Rogers, Zimmerman, and Gavini]{GB2023}
Das,~S.; Bikash,~K.; Subramanian,~V.; Panigrahi,~G.; Motamarri,~P.; Rogers,~D.; Zimmerman,~P.~M.; Gavini,~V. Large-scale materials modeling at quantum accuracy: Ab initio simulations of quasicrystals and interacting extended defects in metallic alloys. In Proceedings of the International Conference for High Performance Computing, Networking, Storage and Analysis (SC'23). 2023; pp 1--12\relax
\mciteBstWouldAddEndPuncttrue
\mciteSetBstMidEndSepPunct{\mcitedefaultmidpunct}
{\mcitedefaultendpunct}{\mcitedefaultseppunct}\relax
\EndOfBibitem
\bibitem[Zhuravel \latin{et~al.}(2020)Zhuravel, Huang, Polycarpou, Polydorides, Motamarri, Katrivas, Rotem, Sperling, Zotti, Kotlyar, \latin{et~al.} others]{zhuravel2020backbone}
Zhuravel,~R.; Huang,~H.; Polycarpou,~G.; Polydorides,~S.; Motamarri,~P.; Katrivas,~L.; Rotem,~D.; Sperling,~J.; Zotti,~L.~A.; Kotlyar,~A.~B.; others Backbone charge transport in double-stranded DNA. \emph{Nature Nanotechnology} \textbf{2020}, \emph{15}, 836--840\relax
\mciteBstWouldAddEndPuncttrue
\mciteSetBstMidEndSepPunct{\mcitedefaultmidpunct}
{\mcitedefaultendpunct}{\mcitedefaultseppunct}\relax
\EndOfBibitem
\bibitem[Kumar \latin{et~al.}(2023)Kumar, Ludhwani, Das, Gavini, Kanjarla, and Adlakha]{kumar2023effect}
Kumar,~P.; Ludhwani,~M.~M.; Das,~S.; Gavini,~V.; Kanjarla,~A.; Adlakha,~I. Effect of hydrogen on plasticity of $\alpha$-Fe: A multi-scale assessment. \emph{International Journal of Plasticity} \textbf{2023}, \emph{165}, 103613\relax
\mciteBstWouldAddEndPuncttrue
\mciteSetBstMidEndSepPunct{\mcitedefaultmidpunct}
{\mcitedefaultendpunct}{\mcitedefaultseppunct}\relax
\EndOfBibitem
\bibitem[Yao \latin{et~al.}(2022)Yao, Das, Liu, Wu, Cheng, Gavini, and Xiao]{yao2022modulating}
Yao,~L.; Das,~S.; Liu,~X.; Wu,~K.; Cheng,~Y.; Gavini,~V.; Xiao,~B. Modulating the microscopic lattice distortions through the Al-rich layers for boosting the ferroelectricity in Al: HfO2 nanofilms. \emph{Journal of Physics D: Applied Physics} \textbf{2022}, \emph{55}, 455501\relax
\mciteBstWouldAddEndPuncttrue
\mciteSetBstMidEndSepPunct{\mcitedefaultmidpunct}
{\mcitedefaultendpunct}{\mcitedefaultseppunct}\relax
\EndOfBibitem
\bibitem[Ghosh \latin{et~al.}(2019)Ghosh, Ma, Gavini, and Galli]{ghosh2019all}
Ghosh,~K.; Ma,~H.; Gavini,~V.; Galli,~G. All-electron density functional calculations for electron and nuclear spin interactions in molecules and solids. \emph{Physical Review Materials} \textbf{2019}, \emph{3}, 043801\relax
\mciteBstWouldAddEndPuncttrue
\mciteSetBstMidEndSepPunct{\mcitedefaultmidpunct}
{\mcitedefaultendpunct}{\mcitedefaultseppunct}\relax
\EndOfBibitem
\bibitem[Krishnendu \latin{et~al.}(2021)Krishnendu, He, Mykyta, Vikram, and Giulia]{krishnendu2021spin}
Krishnendu,~G.; He,~M.; Mykyta,~O.; Vikram,~G.; Giulia,~G. Spin--spin interactions in defects in solids from mixed all-electron and pseudopotential first-principles calculations. \emph{NPJ Computational Materials} \textbf{2021}, \emph{7}\relax
\mciteBstWouldAddEndPuncttrue
\mciteSetBstMidEndSepPunct{\mcitedefaultmidpunct}
{\mcitedefaultendpunct}{\mcitedefaultseppunct}\relax
\EndOfBibitem
\bibitem[Kanungo \latin{et~al.}(2019)Kanungo, Zimmerman, and Gavini]{kanungo2019exact}
Kanungo,~B.; Zimmerman,~P.~M.; Gavini,~V. Exact exchange-correlation potentials from ground-state electron densities. \emph{Nature Communications} \textbf{2019}, \emph{10}, 1--9\relax
\mciteBstWouldAddEndPuncttrue
\mciteSetBstMidEndSepPunct{\mcitedefaultmidpunct}
{\mcitedefaultendpunct}{\mcitedefaultseppunct}\relax
\EndOfBibitem
\bibitem[Kanungo \latin{et~al.}(2021)Kanungo, Zimmerman, and Gavini]{kanungo2021comparison}
Kanungo,~B.; Zimmerman,~P.~M.; Gavini,~V. A comparison of exact and model exchange--correlation potentials for molecules. \emph{The Journal of Physical Chemistry Letters} \textbf{2021}, \emph{12}, 12012--12019\relax
\mciteBstWouldAddEndPuncttrue
\mciteSetBstMidEndSepPunct{\mcitedefaultmidpunct}
{\mcitedefaultendpunct}{\mcitedefaultseppunct}\relax
\EndOfBibitem
\bibitem[Kanungo \latin{et~al.}(2023)Kanungo, Hatch, Zimmerman, and Gavini]{kanungo2023}
Kanungo,~B.; Hatch,~J.; Zimmerman,~P.~M.; Gavini,~V. Exact and model exchange--correlation potentials for open shell systems. \emph{The Journal of Physical Chemistry Letters} \textbf{2023}, \emph{14}, 10039--10045\relax
\mciteBstWouldAddEndPuncttrue
\mciteSetBstMidEndSepPunct{\mcitedefaultmidpunct}
{\mcitedefaultendpunct}{\mcitedefaultseppunct}\relax
\EndOfBibitem
\bibitem[Khoromskaia and Khoromskij(2015)Khoromskaia, and Khoromskij]{khoromskaia2015tensor}
Khoromskaia,~V.; Khoromskij,~B.~N. Tensor numerical methods in quantum chemistry: from Hartree--Fock to excitation energies. \emph{Physical Chemistry Chemical Physics} \textbf{2015}, \emph{17}, 31491--31509\relax
\mciteBstWouldAddEndPuncttrue
\mciteSetBstMidEndSepPunct{\mcitedefaultmidpunct}
{\mcitedefaultendpunct}{\mcitedefaultseppunct}\relax
\EndOfBibitem
\bibitem[Khoromskij and Khoromskaia(2007)Khoromskij, and Khoromskaia]{khoromskij2007low}
Khoromskij,~B.; Khoromskaia,~V. Low rank Tucker-type tensor approximation to classical potentials. \emph{Open Mathematics} \textbf{2007}, \emph{5}, 523--550\relax
\mciteBstWouldAddEndPuncttrue
\mciteSetBstMidEndSepPunct{\mcitedefaultmidpunct}
{\mcitedefaultendpunct}{\mcitedefaultseppunct}\relax
\EndOfBibitem
\bibitem[Khoromskij \latin{et~al.}(2009)Khoromskij, Khoromskaia, Chinnamsetty, and Flad]{khoromskij2009tensor}
Khoromskij,~B.~N.; Khoromskaia,~V.; Chinnamsetty,~S.~R.; Flad,~H.-J. Tensor decomposition in electronic structure calculations on 3D Cartesian grids. \emph{Journal of computational physics} \textbf{2009}, \emph{228}, 5749--5762\relax
\mciteBstWouldAddEndPuncttrue
\mciteSetBstMidEndSepPunct{\mcitedefaultmidpunct}
{\mcitedefaultendpunct}{\mcitedefaultseppunct}\relax
\EndOfBibitem
\bibitem[Braess and Hackbusch(2009)Braess, and Hackbusch]{braess2009efficient}
Braess,~D.; Hackbusch,~W. \emph{Multiscale, Nonlinear and Adaptive Approximation: Dedicated to Wolfgang Dahmen on the Occasion of his 60th Birthday}; Springer, 2009; pp 39--74\relax
\mciteBstWouldAddEndPuncttrue
\mciteSetBstMidEndSepPunct{\mcitedefaultmidpunct}
{\mcitedefaultendpunct}{\mcitedefaultseppunct}\relax
\EndOfBibitem
\bibitem[Giannozzi \latin{et~al.}(2017)Giannozzi, Andreussi, Brumme, Bunau, Nardelli, Calandra, Car, Cavazzoni, Ceresoli, Cococcioni, \latin{et~al.} others]{giannozzi2017advanced}
Giannozzi,~P.; Andreussi,~O.; Brumme,~T.; Bunau,~O.; Nardelli,~M.~B.; Calandra,~M.; Car,~R.; Cavazzoni,~C.; Ceresoli,~D.; Cococcioni,~M.; others Advanced capabilities for materials modelling with Quantum ESPRESSO. \emph{Journal of physics: Condensed matter} \textbf{2017}, \emph{29}, 465901\relax
\mciteBstWouldAddEndPuncttrue
\mciteSetBstMidEndSepPunct{\mcitedefaultmidpunct}
{\mcitedefaultendpunct}{\mcitedefaultseppunct}\relax
\EndOfBibitem
\bibitem[Levy(1979)]{levy1979universal}
Levy,~M. Universal variational functionals of electron densities, first-order density matrices, and natural spin-orbitals and solution of the v-representability problem. \emph{Proceedings of the National Academy of Sciences} \textbf{1979}, \emph{76}, 6062--6065\relax
\mciteBstWouldAddEndPuncttrue
\mciteSetBstMidEndSepPunct{\mcitedefaultmidpunct}
{\mcitedefaultendpunct}{\mcitedefaultseppunct}\relax
\EndOfBibitem
\bibitem[Garrick \latin{et~al.}(2020)Garrick, Natan, Gould, and Kronik]{garrick2020exact}
Garrick,~R.; Natan,~A.; Gould,~T.; Kronik,~L. Exact Generalized Kohn-Sham Theory for Hybrid Functionals. \emph{Physical Review X} \textbf{2020}, \emph{10}, 021040\relax
\mciteBstWouldAddEndPuncttrue
\mciteSetBstMidEndSepPunct{\mcitedefaultmidpunct}
{\mcitedefaultendpunct}{\mcitedefaultseppunct}\relax
\EndOfBibitem
\bibitem[Ren \latin{et~al.}(2012)Ren, Rinke, Blum, Wieferink, Tkatchenko, Sanfilippo, Reuter, and Scheffler]{ren2012resolution}
Ren,~X.; Rinke,~P.; Blum,~V.; Wieferink,~J.; Tkatchenko,~A.; Sanfilippo,~A.; Reuter,~K.; Scheffler,~M. Resolution-of-identity approach to Hartree--Fock, hybrid density functionals, RPA, MP2 and GW with numeric atom-centered orbital basis functions. \emph{New Journal of Physics} \textbf{2012}, \emph{14}, 053020\relax
\mciteBstWouldAddEndPuncttrue
\mciteSetBstMidEndSepPunct{\mcitedefaultmidpunct}
{\mcitedefaultendpunct}{\mcitedefaultseppunct}\relax
\EndOfBibitem
\bibitem[Levchenko \latin{et~al.}(2015)Levchenko, Ren, Wieferink, Johanni, Rinke, Blum, and Scheffler]{levchenko2015hybrid}
Levchenko,~S.~V.; Ren,~X.; Wieferink,~J.; Johanni,~R.; Rinke,~P.; Blum,~V.; Scheffler,~M. Hybrid functionals for large periodic systems in an all-electron, numeric atom-centered basis framework. \emph{Computer Physics Communications} \textbf{2015}, \emph{192}, 60--69\relax
\mciteBstWouldAddEndPuncttrue
\mciteSetBstMidEndSepPunct{\mcitedefaultmidpunct}
{\mcitedefaultendpunct}{\mcitedefaultseppunct}\relax
\EndOfBibitem
\bibitem[Anderson(1965)]{anderson1965iterative}
Anderson,~D.~G. Iterative procedures for nonlinear integral equations. \emph{Journal of the ACM (JACM)} \textbf{1965}, \emph{12}, 547--560\relax
\mciteBstWouldAddEndPuncttrue
\mciteSetBstMidEndSepPunct{\mcitedefaultmidpunct}
{\mcitedefaultendpunct}{\mcitedefaultseppunct}\relax
\EndOfBibitem
\bibitem[Zhou \latin{et~al.}(2006)Zhou, Saad, Tiago, and Chelikowsky]{zhou2006self}
Zhou,~Y.; Saad,~Y.; Tiago,~M.~L.; Chelikowsky,~J.~R. Self-consistent-field calculations using Chebyshev-filtered subspace iteration. \emph{Journal of Computational Physics} \textbf{2006}, \emph{219}, 172--184\relax
\mciteBstWouldAddEndPuncttrue
\mciteSetBstMidEndSepPunct{\mcitedefaultmidpunct}
{\mcitedefaultendpunct}{\mcitedefaultseppunct}\relax
\EndOfBibitem
\bibitem[Hu \latin{et~al.}(2017)Hu, Lin, Banerjee, Vecharynski, and Yang]{hu2017adaptively}
Hu,~W.; Lin,~L.; Banerjee,~A.~S.; Vecharynski,~E.; Yang,~C. Adaptively compressed exchange operator for large-scale hybrid density functional calculations with applications to the adsorption of water on silicene. \emph{Journal of chemical theory and computation} \textbf{2017}, \emph{13}, 1188--1198\relax
\mciteBstWouldAddEndPuncttrue
\mciteSetBstMidEndSepPunct{\mcitedefaultmidpunct}
{\mcitedefaultendpunct}{\mcitedefaultseppunct}\relax
\EndOfBibitem
\bibitem[Kolda and Bader(2009)Kolda, and Bader]{kolda2009tensor}
Kolda,~T.~G.; Bader,~B.~W. Tensor decompositions and applications. \emph{SIAM review} \textbf{2009}, \emph{51}, 455--500\relax
\mciteBstWouldAddEndPuncttrue
\mciteSetBstMidEndSepPunct{\mcitedefaultmidpunct}
{\mcitedefaultendpunct}{\mcitedefaultseppunct}\relax
\EndOfBibitem
\bibitem[Khoromskij \latin{et~al.}(2009)Khoromskij, Khoromskaia, Chinnamsetty, and Flad]{Khoromskij2009}
Khoromskij,~B.; Khoromskaia,~V.; Chinnamsetty,~S.; Flad,~H.-J. Tensor decomposition in electronic structure calculations on 3D Cartesian grids. \emph{Journal of Computational Physics} \textbf{2009}, \emph{228}, 5749--5762\relax
\mciteBstWouldAddEndPuncttrue
\mciteSetBstMidEndSepPunct{\mcitedefaultmidpunct}
{\mcitedefaultendpunct}{\mcitedefaultseppunct}\relax
\EndOfBibitem
\bibitem[Blesgen \latin{et~al.}(2012)Blesgen, Gavini, and Khoromskaia]{Blesgen2012}
Blesgen,~T.; Gavini,~V.; Khoromskaia,~V. Approximation of the electron density of Aluminium clusters in tensor-product format. \emph{Journal of Computational Physics} \textbf{2012}, \emph{231}, 2551--2564\relax
\mciteBstWouldAddEndPuncttrue
\mciteSetBstMidEndSepPunct{\mcitedefaultmidpunct}
{\mcitedefaultendpunct}{\mcitedefaultseppunct}\relax
\EndOfBibitem
\bibitem[Lin \latin{et~al.}(2021)Lin, Motamarri, and Gavini]{Lin2021}
Lin,~C.-C.; Motamarri,~P.; Gavini,~V. Tensor-structured algorithm for reduced-order scaling large-scale Kohn–Sham density functional theory calculations. \emph{npj Computational Materials} \textbf{2021}, \emph{7}, 50\relax
\mciteBstWouldAddEndPuncttrue
\mciteSetBstMidEndSepPunct{\mcitedefaultmidpunct}
{\mcitedefaultendpunct}{\mcitedefaultseppunct}\relax
\EndOfBibitem
\bibitem[Motamarri \latin{et~al.}(2016)Motamarri, Gavini, and Blesgen]{Motamarri2016}
Motamarri,~P.; Gavini,~V.; Blesgen,~T. Tucker-tensor algorithm for large-scale Kohn-Sham density functional theory calculations. \emph{Phys. Rev. B} \textbf{2016}, \emph{93}, 125104\relax
\mciteBstWouldAddEndPuncttrue
\mciteSetBstMidEndSepPunct{\mcitedefaultmidpunct}
{\mcitedefaultendpunct}{\mcitedefaultseppunct}\relax
\EndOfBibitem
\bibitem[Hughes(2012)]{hughes2012finite}
Hughes,~T.~J. \emph{The finite element method: linear static and dynamic finite element analysis}; Courier Corporation, 2012\relax
\mciteBstWouldAddEndPuncttrue
\mciteSetBstMidEndSepPunct{\mcitedefaultmidpunct}
{\mcitedefaultendpunct}{\mcitedefaultseppunct}\relax
\EndOfBibitem
\bibitem[Tsuchida and Tsukada(1995)Tsuchida, and Tsukada]{tsuchida1995electronic}
Tsuchida,~E.; Tsukada,~M. Electronic-structure calculations based on the finite-element method. \emph{Physical Review B} \textbf{1995}, \emph{52}, 5573\relax
\mciteBstWouldAddEndPuncttrue
\mciteSetBstMidEndSepPunct{\mcitedefaultmidpunct}
{\mcitedefaultendpunct}{\mcitedefaultseppunct}\relax
\EndOfBibitem
\bibitem[Tsuchida and Tsukada(1996)Tsuchida, and Tsukada]{tsuchida1996adaptive}
Tsuchida,~E.; Tsukada,~M. Adaptive finite-element method for electronic-structure calculations. \emph{Physical Review B} \textbf{1996}, \emph{54}, 7602\relax
\mciteBstWouldAddEndPuncttrue
\mciteSetBstMidEndSepPunct{\mcitedefaultmidpunct}
{\mcitedefaultendpunct}{\mcitedefaultseppunct}\relax
\EndOfBibitem
\bibitem[Ballard \latin{et~al.}(2020)Ballard, Klinvex, and Kolda]{ballard2020tuckermpi}
Ballard,~G.; Klinvex,~A.; Kolda,~T.~G. TuckerMPI: A parallel C++/MPI software package for large-scale data compression via the tucker tensor decomposition. \emph{ACM Transactions on Mathematical Software (TOMS)} \textbf{2020}, \emph{46}, 1--31\relax
\mciteBstWouldAddEndPuncttrue
\mciteSetBstMidEndSepPunct{\mcitedefaultmidpunct}
{\mcitedefaultendpunct}{\mcitedefaultseppunct}\relax
\EndOfBibitem
\bibitem[Hamann(2013)]{PhysRevB.88.085117}
Hamann,~D.~R. Optimized norm-conserving Vanderbilt pseudopotentials. \emph{Phys. Rev. B} \textbf{2013}, \emph{88}, 085117\relax
\mciteBstWouldAddEndPuncttrue
\mciteSetBstMidEndSepPunct{\mcitedefaultmidpunct}
{\mcitedefaultendpunct}{\mcitedefaultseppunct}\relax
\EndOfBibitem
\end{mcitethebibliography}
	
\end{document}